\documentclass[aps, prl,twocolumn, superscriptaddress, preprintnumbers, floatfix, longbibliography, 10pt]{revtex4-2}

\usepackage{graphicx, xcolor}
\usepackage{amsfonts,amsmath,amssymb,amstext,dsfont,bm,mathtools}
\usepackage{float,wrapfig,subfigure,psfrag}
\usepackage{comment}

\usepackage[colorlinks=true,urlcolor=blue,citecolor=blue,linkcolor=blue,breaklinks=true]{hyperref}

\newcommand{\be}{\begin{equation}}
\newcommand{\ee}{\end{equation}}
\newcommand{\ba}{\begin{eqnarray}}
\newcommand{\ea}{\end{eqnarray}}

\definecolor{purple}{rgb}{0.8,0,0.6}
\definecolor{darkgreen}{rgb}{0.00,0.6,0.00}

\definecolor{Blue}{rgb}{0,0,0.85}

\extrafloats{100}

\begin{document}

\title{Electric and spin current vortices in altermagnets}
\date{February 17, 2026}

\author{Arsen~Herasymchuk}
\email{arsengerasymchuk@gmail.com}
\affiliation{Bogolyubov Institute for Theoretical Physics, Kyiv, 03143, Ukraine}
%\affiliation{Department of Physics, Taras Shevchenko National Kyiv University, Kyiv, 01601, Ukraine}

\author{Karl Bergson Hallberg}
\affiliation{Center for Quantum Spintronics, Department of Physics, Norwegian University of Science and Technology, NO-7491 Trondheim, Norway}

\author{Erik Wegner Hodt}
\affiliation{Center for Quantum Spintronics, Department of Physics, Norwegian University of Science and Technology, NO-7491 Trondheim, Norway}

\author{Jacob Linder}
\affiliation{Center for Quantum Spintronics, Department of Physics, Norwegian University of Science and Technology, NO-7491 Trondheim, Norway}

\author{E.~V.~Gorbar}
\email{gorbar@knu.ua}
\affiliation{Department of Physics, Taras Shevchenko National Kyiv University, Kyiv, 01601, Ukraine}
\affiliation{Bogolyubov Institute for Theoretical Physics, Kyiv, 03143, Ukraine}

\author{Pavlo~Sukhachov}
\email{pavlo.sukhachov@missouri.edu}
\affiliation{Department of Physics and Astronomy, University of Missouri, Columbia, Missouri, 65211, USA}
\affiliation{MU Materials Science \& Engineering Institute, University of Missouri, Columbia, Missouri, 65211, USA}

\begin{abstract}
Altermagnets constitute a class of collinear magnets with momentum-dependent spin splitting and vanishing net magnetization. Direct observation of the characteristic altermagnetic spin splitting, however, remains challenging. Indirect signatures can be obtained via transport studies, which so far have only considered homogeneous driving fields.
%fields driving currents.
We propose to leverage nonuniform electric fields and spin density gradients to probe the shape and the spin polarization of altermagnetic Fermi surfaces via transport measurements. By using both a semiclassical Boltzmann approach and a lattice Keldysh formalism, we show that altermagnets excite swirling electric and spin currents whose profiles depend on the relative orientation of altermagnetic lobes with respect to the sample boundaries. These currents can be measured via magnetometry techniques. Unlike previous proposals considering the hydrodynamic regime of transport, swirling currents are observed even in the Ohmic regime and rely exclusively on the altermagnetic spin splitting, with no swirls observed in ferromagnets. The electric and spin current vortices predicted here provide a different altermagnetic signature in an experimentally accessible setup.
\end{abstract}

\maketitle

\textit{Introduction.}
Transport of charge and spin represents one of the fundamental types of measurements in condensed matter physics that is indispensable in identifying and investigating new materials. Among such materials that have recently attracted much attention are magnetic materials with anisotropic nonrelativistically spin-split energy bands~\cite{Noda-Nakamura-MomentumdependentBandSpin-2016, Smejkal-Sinova:2020, Hayami-Kusunose-MomentumDependentSpinSplitting-2019, Ahn-Kunes:2019, Yuan-Zunger:2020, Yuan-Zunger-PredictionLowZCollinear-2021, Ma-Liu-MultifunctionalAntiferromagneticMaterials-2021, Smejkal-Jungwirth:2022, Smejkal-Jungwirth-ConventionalFerromagnetismAntiferromagnetism-2022, Mazin:2022}. Known as altermagnets~\cite{Smejkal-Sinova:2020, Smejkal-Jungwirth:2022}, such materials are characterized by a combined symmetry including lattice rotations and spin reversal. Altermagnets preserve the inversion symmetry but break the time-reversal symmetry (TRS).

Altermagnetism was predicted in several materials such as MnTe, a few atom layers of RuO$_2$, Mn$_5$Si$_3$, MnF$_2$, V$_2$Se$_2$O, V$_2$Te$_2$O, CrO, CrSb, and intercalated transition metal dichalcogenides CoNb$_4$Se$_8$, and $\kappa$-type organic antiferromagnets~\cite{Smejkal-Jungwirth:2022, Bai-Yao-AltermagnetismExploringNew-2024, Naka-Seo-AltermagneticPerovskites-2025}. Signatures of spin-split energy bands were observed in angle-resolved photoemission spectroscopy (ARPES)~\cite{Fedchenko-Elmers-ObservationTimereversalSymmetry-2024, Krempasky-Jungwirth:2024, Lee-Kim-BrokenKramersDegeneracy-2024, Osumi-Sato-ObservationGiantBand-2024, Zeng-Liu-ObservationSpinSplitting-2024, Ding-Shen-LargeBandsplittingWave-2024, Jeong-Jalan-AltermagneticPolarMetallic-2024, Dale-Griffin-NonrelativisticSpinSplitting-2024, Zhang-Chen-CrystalsymmetrypairedSpinValley-2025, Jiang-Qian-MetallicRoomtemperatureDwave-2025} and circular dichroism~\cite{Hariki2023}. However, ARPES measurements generically suffer from low resolution and strong reliance on the comparison between the experimental data and the \textit{ab initio} calculations.

Transport experiments in altermagnets are mostly focused on the observation of the anomalous Hall effect (AHE)~\cite{GonzalezBetancourt-Kriegner-SpontaneousAnomalousHall-2023, Reichlova-Smejkal-ObservationSpontaneousAnomalous-2024, Leiviska-Baltz-AnisotropyAnomalousHall-2024}; see also Refs.~\cite{Solovyev-MagnetoopticalEffectWeak-1997, Naka-Seo-AnomalousHallEffect-2020} for earlier discussions of the AHE in altermagnetic candidates. The AHE, however, may originate from different sources, such as a reduced symmetry at the interface~\cite{Shao-Tsymbal-InterfacialCrystalHall-2021}, domain walls~\cite{Xia-Guo-GiantDomainWall-2024}, or spin canting~\cite{Dzyaloshinsky-ThermodynamicTheoryWeak-1958, Moriya-AnisotropicSuperexchangeInteraction-1960}, see also Ref.~\cite{Galindez-Ruales-Klaui-AltermagnetismHoppingRegime-2024}. Therefore, the AHE alone cannot be used to unambiguously prove the altermagnetic nature of a material.
The spin-splitter effect~\cite{Naka-Seo-SpinCurrentGeneration-2019, Gonzalez-Hernandez-Zelezny:2021, Karube-Nitta:2022}, which is the generation of the spin current by an applied electric field, suggests a different alternative to the AHE for altermagnets. However, its observed signatures are limited to spin-splitter torque in RuO$_2$~\cite{Karube-Nitta:2022}, whose magnetic state is still debated.
Thus, other types of probes are beneficial to rigorously prove the realization of the altermagnetic state.

In this Letter, we propose to use point contacts in altermagnets to generate swirling electric and spin currents whose patterns are directly linked to the symmetry of the altermagnetic spin splitting and the geometry of the sample. While vortices in electron fluids are usually considered as a key indicator of electron hydrodynamics~\cite{Nazaryan-Levitov-NonlocalConductivityContinued-2024}, swirling currents in our work appear in the Ohmic transport regime. The proposed effects can be readily investigated via available experimental methods. In particular, the predicted swirling electric currents can be probed via quantum spin magnetometry~\cite{Levine-Walsworth-PrinciplesTechniquesQuantum-2019, Barry-Walsworth:2019-QSM}, whereas the distribution of the electrochemical potential can be measured via scanning tunneling potentiometry~\cite{Baddorf-ScanningTunnelingPotentiometry-2007}. Local spin currents can also be probed via techniques similar to the non-local spin valve measurement~\cite{Johnson-Silsbee-InterfacialChargespinCoupling-1985, Jedema-VanWees-ElectricalSpinInjection-2001}. Thus, our results complement previous transport measurements and provide an opportunity to cross-verify electric and spin currents via a different set of techniques.

\textit{Continuum model and kinetic equations.}
We start with defining the low-energy continuum model of altermagnets. We focus on two-dimensional (2D) altermagnets as they provide the most direct way to visualize electric and spin currents; the generalization of our formalism to three-dimensional cases is straightforward.

We employ the following model of an altermagnet with a momentum-dependent Zeeman field~\footnote{As we verify in Sec.~S1.B.2 (see Supplemental Material~\cite{SM}), a model with sublattice degrees of freedom leads to the same structure of the response tensor.}:
\begin{equation}
\label{eq:hamiltonian}
H=t_0 k^2 +\sigma_z J(\mathbf{k}) - \hat{\mu}.
\end{equation}
Here, $\mathbf{k}$ is the momentum vector, $k=\sqrt{k_x^2+k_y^2}$, $\sigma_{z}$ is the Pauli matrix in the spin space, $J(\mathbf{k})$ is an effective exchange field, $\hat{\mu}=\text{diag} \left( \mu_{+}, \mu_{-} \right)$, where $\mu_{\lambda}$ is the spin-resolved chemical potential for the spin projection $\lambda=\pm$~\footnote{The spin-dependent effective chemical potential can be achieved by applying a weak magnetic field via the Zeeman term. In 2D systems, orbital effects can be ignored if the field is in the plane of the material.}, and parameter $t_0$ corresponds to the inverse effective mass of quasiparticles. To obtain finite Fermi surfaces, we assume that $t_0k^2 > \left|J(\mathbf{k})\right|$.

\begin{figure}[t]
\includegraphics[width=.4\textwidth]{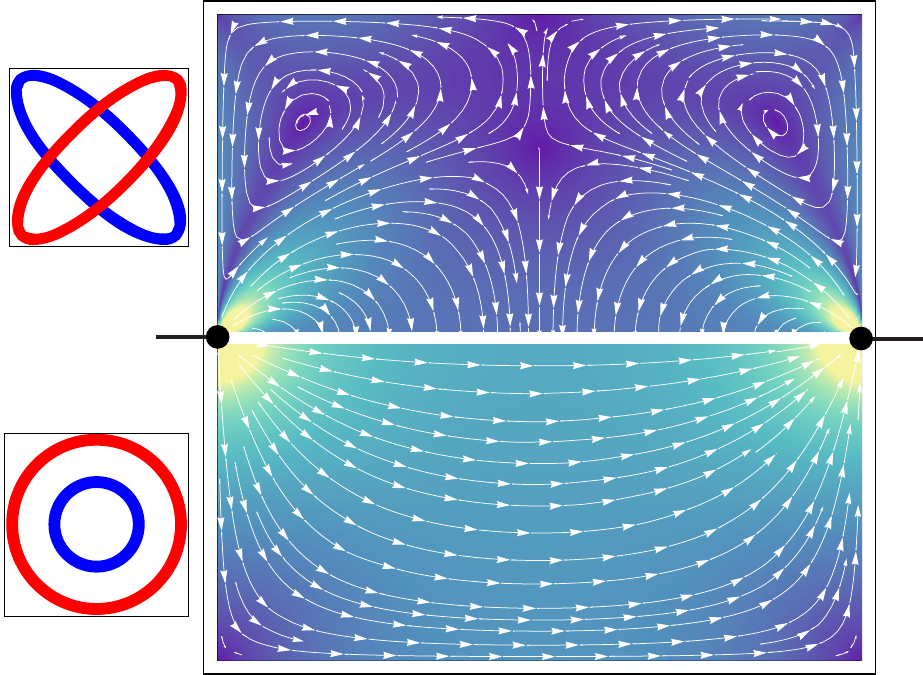}
\caption{
Schematic setup of a magnet with point-like contacts. The streamlines illustrate the flow of electric (spin) currents induced by the applied spin density (electric potential) difference at the contacts. The upper half corresponds to a $d$-wave altermagnet and the lower half represents a ferromagnet; the corresponding Fermi surfaces with spin-up and spin-down polarized bands are shown in red and blue. Despite the system being diffusive and non-interacting, current vortices arise in the altermagnetic case, unlike the ferromagnetic one.
}
\label{fig:band}
\end{figure}

The dispersion relation of electron quasiparticles is
\begin{equation}
\label{eq:spectra}
\varepsilon_{\lambda}=t_0 k^2 +\lambda J(\mathbf{k}).
\end{equation}
The corresponding Fermi surfaces for a $d$-wave altermagnet are schematically shown by the red ($\lambda=+$) and blue ($\lambda=-$) ellipses in Fig.~\ref{fig:band}.

To address the transport properties of the continuum model, we employ the standard semiclassical approach in which the dynamics of quasiparticles is determined by the Boltzmann kinetic equation~\footnote{A quantum transport theory for generic magnetic metals, valid both for normal and superconducting states, was recently developed in Ref.~\cite{Kokkeler-Bergeret-QuantumTransportTheory-2025} based on the generalization of nonlinear sigma model},
\begin{equation}
\label{eq:Boltzmann-def}
\partial_t f_{\lambda} +\left( \mathbf{v}_{\mathbf{k},\lambda} \cdot \bm{\nabla}f_{\lambda}\right)+ \left( e \mathbf{E}\cdot \partial_\mathbf{k} f_{\lambda} \right) = I_{\text{col}}\left\{ f_{\lambda} \right\}.
\end{equation}
Here $f_{\lambda}=f_{\lambda}(t,\mathbf{r},\mathbf{k})$ is the distribution function of quasiparticles wit the spin $\lambda$, $\mathbf{v}_{\mathbf{k},\lambda} = \partial_{\mathbf{k}} \varepsilon_{\mathbf{\lambda}}$ is the quasiparticle velocity, $e<0$ is the electron charge, $\mathbf{E}$ is an electric field, and $I_{\text{col}}\left\{ f_{\lambda} \right\}$ is the collision integral which describes scattering off disorder.

For simplicity, we assume nonmagnetic disorder and use the standard relaxation time approximation $I_{\text{col}}\left\{ f_{\lambda} \right\} =-\left(f_{\lambda}- \langle f_{\lambda}\rangle\right)/\tau$, where $\tau$ is the relaxation time and $\langle f_{\lambda}\rangle$ is the averaged over the Fermi-surface distribution function. Since we assumed nonmagnetic elastic impurity scattering, the spin-flip processes are determined by the spin-orbital coupling (see, e.g., Ref.~\cite{Matsuo:2017}) and, therefore, are expected to be weak in altermagnets. Weak spin-flip processes reduce the magnitude of the current but do not qualitatively change the current distribution (see the Supplementary Material~\cite{SM}).

The charge and current densities in a 2D system for quasiparticles with the spin projection $\lambda$ are
\begin{equation}
\label{eq:Boltzmann-rho-j-def}
\left\{\rho_\lambda (t,\mathbf{r}), \mathbf{j}_\lambda (t,\mathbf{r})\right\} = e\int\frac{d^2k}{(2\pi)^2} \left\{1, \mathbf{v}_{\mathbf{k},\lambda}\right\} f_{\lambda} (t,\mathbf{r},\mathbf{k}).
\end{equation}
Electric and spin currents are obtained as $\mathbf{j}_{\rm el}(t,\mathbf{r}) = \sum_{\lambda} \mathbf{j}_\lambda (t,\mathbf{r})$ and $\mathbf{j}_{\rm sp}(t,\mathbf{r}) = \sum_{\lambda} \lambda\, \mathbf{j}_\lambda (t,\mathbf{r})$.

By integrating the Boltzmann equation \eqref{eq:Boltzmann-def} over momenta, we derive the continuity equation
\begin{equation}
\label{eq:rho-eq}
\partial_t \rho_\lambda+\left(\bm{\nabla} \cdot  \mathbf{j}_\lambda \right)=0.
\end{equation}
Multiplying Eq.~\eqref{eq:Boltzmann-def} by $e\mathbf{v}_{\mathbf{k},\lambda}$ and integrating over momenta, we obtain
\begin{equation}
\label{eq:j-eq}
\partial_t j_{\lambda,j} + \nabla_{i} \tilde{\Pi}_{ij} +e\Pi_{ij}E_i  = - \frac{j_{\lambda,j}}{\tau},
\end{equation}
where
\begin{eqnarray}
\label{eq:Pi-ij}
\tilde{\Pi}_{ij}(t,\mathbf{r}) &=& e\int\frac{d^2k}{(2\pi)^2} v_{\mathbf{k},\lambda,i} v_{\mathbf{k},\lambda,j} f_{\lambda} (t,\mathbf{r},\mathbf{k}),\\
\label{eq:tPi-ij}
\Pi_{ij}(t,\mathbf{r}) &=&
-e\int\frac{d^2k}{(2\pi)^2} (\partial_{k_i} \partial_{k_j} \varepsilon_{\lambda}) f_{\lambda} (t,\mathbf{r},\mathbf{k}).
\end{eqnarray}
Assuming weak deviations from equilibrium, neglecting temperature deviations, and linearizing Eq.~\eqref{eq:j-eq}, we derive the following equation for the current density:
\begin{equation}
\label{eq:j-eq-2}
\partial_t j_{\lambda,l}  +\frac{\sigma_{il}}{\tau} \nabla_{i} \bar{\phi}_{\lambda}  = - \frac{j_{\lambda,l}}{\tau}.
\end{equation}
In writing this equation, we introduced the conductivity tensor
\begin{equation}
\label{eq:Pi-0-ij}
\sigma_{ij} = 2 e \tau \rho^{(0)}_{\lambda} \left( t_0  \delta_{ij} + \lambda T_{\lambda, ij} \right),
\end{equation}
where
\begin{equation}
\label{eq:T-ij}
T_{\lambda, ij} =\frac{e}{2 \rho^{(0)}_{\lambda} } \int\frac{d^2k}{(2\pi)^2} \left[ \partial_{k_i} \partial_{k_j} J(\mathbf{k}) \right] f_{\lambda}^{(0)}(\mathbf{k})
\end{equation}
is the part of the conductivity tensor determined by the effective exchange field, $f^{(0)}_{\lambda} (\mathbf{k}) =1/\left[e^{\left(\varepsilon_{\lambda} -\mu_{\lambda} \right)/T}+1\right]$ is the equilibrium Fermi-Dirac distribution function, $\rho^{(0)}_{\lambda}$ is the equilibrium charge density, and $e\bar{\phi}_{\lambda} = \delta \mu_{\lambda} +e\phi$ is the effective spin-resolved electrochemical potential with $\phi$ being electric potential and $\delta \mu_{\lambda}$ being the spin-resolved deviation of the chemical potential from its equilibrium value. The off-diagonal and spin-dependent structure of the conductivity tensor in Eq.~\eqref{eq:Pi-0-ij} originates from altermagnetic spin splitting and agrees with that reported in Ref.~\cite{Kokkeler-Bergeret-QuantumTransportTheory-2025}.

Equation \eqref{eq:j-eq-2} is general and can be applied both for static and dynamic perturbations. In this Letter, we focus on the stationary transport regime, where Eq.~\eqref{eq:j-eq-2} reduces to
\begin{equation}
\label{eq:nonlocal-eq-fin}
\nabla_{j} \left(\sigma_{ij} \nabla_{i}\bar{\phi}_{\lambda} \right)=0.
\end{equation}
This differential equation should be supplemented with boundary conditions. For example, one can fix $\bar{\phi}_{\lambda}$ at the contacts or require a constant injected current. Then, Eq.~\eqref{eq:nonlocal-eq-fin} is a partial differential equation that can be solved numerically or, in certain cases, such as a ribbon with the fixed values of current at the point-like contacts (see the Supplemental Material (SM)~\cite{SM}), analytically.

\textit{Vorticity in the Ohmic transport regime.}
Before delving into the numerical solutions to Eq.~\eqref{eq:nonlocal-eq-fin}, let us present intuitive arguments supporting the possibility of vortices in the Ohmic transport regime. We start with the analog of the vorticity equation in fluid mechanics~\cite{Landau:t6-2013} and consider the spin-resolved analog of vorticity $\bm{\omega}_{\lambda} = \bm{\nabla}\times \mathbf{j}_{\lambda}$. Assuming the absence of the equilibrium spin density, $\mu_{\lambda}=\mu$, the equations for the electric $\bm{\omega}_{\rm el} = \sum_{\lambda}\bm{\omega}_{\lambda}$ and spin $\bm{\omega}_{\rm sp} = \sum_{\lambda}\lambda\bm{\omega}_{\lambda}$ vorticities follow from Eq.~\eqref{eq:j-eq-2}
\begin{eqnarray}
\label{eq:vorticity-eq-el}
&&\partial_t \omega_{{\rm el}, j} +\frac{\omega_{{\rm el}, j}}{\tau} = -2e \rho^{(0)} \epsilon_{jlm} T_{im}\nabla_l \nabla_i \bar{\phi}_{\rm sp},\\
\label{eq:vorticity-eq-sp}
&&\partial_t \omega_{{\rm sp}, j} +\frac{\omega_{{\rm sp}, j}}{\tau} = -2e \rho^{(0)} \epsilon_{jlm} T_{im}\nabla_l \nabla_i \bar{\phi}_{\rm el},
\end{eqnarray}
where $\epsilon_{jlm}$ is the fully antisymmetric tensor. These equations allow us to identify two sources of vorticity: (i) off-diagonal elements of the altermagnetic part of the conductivity tensor $T_{\lambda, i\neq m}$ and (ii) anisotropy in the diagonal elements $T_{\lambda, ii}\neq T_{\lambda, mm}$ with $i\neq m$. The former allows for the vorticity even when potentials change only along one direction, e.g., for rectangular contacts. The off-diagonal terms can originate, e.g., from the Hall effect or, as we show in this Letter, naturally appear in altermagnets due to momentum-dependent spin splitting, see Eq.~\eqref{eq:T-ij}. Vorticity for different diagonal elements, which also appear in altermagnets, requires nonuniform potential with $\nabla_l \nabla_i \bar{\phi}_{\lambda} \neq0$ at $l\neq i$. The latter occurs at the edges of rectangular contacts or for point-like contacts. In both cases, nontrivial spatial distribution of the electric potential (spin density) acts as a source of the spin (electric) current vorticity. Therefore, by applying a voltage difference to the contacts, a circulating spin current is expected. In what follows, we confirm this via numerical calculations using two complementary techniques for the semiclassical and fully quantum regimes, respectively.

\begin{figure*}[t]
\centering
\includegraphics[width=1\textwidth]{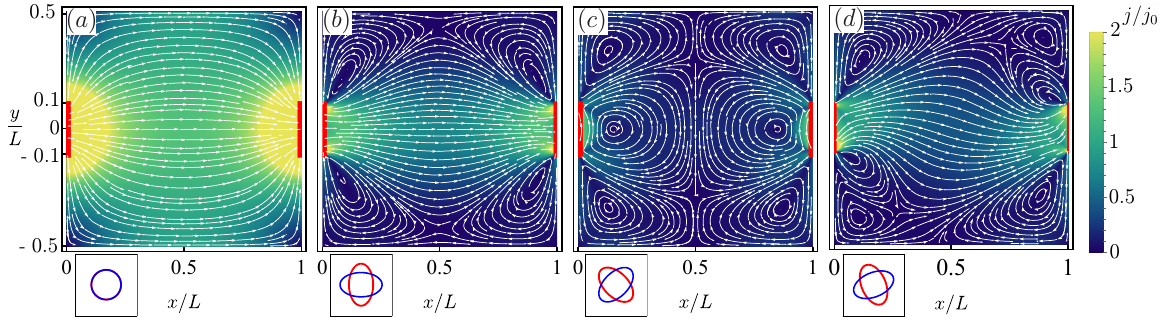}
\caption{
(a) Electric current $j_{\rm el} = j_{\uparrow}+j_{\downarrow}$ distribution without altermagnetism, $t_1=0$ and $t_2=0$. There are no qualitative changes in the shape of the electric current streamlines at nonzero $t_1$ or $t_2$. Spin current $j_{\rm sp} = j_{\uparrow}-j_{\downarrow}$ distribution in altermagnets for (b) $t_1=t_0/2$ and $t_2=0$, (c) $t_1=0$ and $t_2=t_0/2$, and (d) $t_1=t_2=t_0/(2\sqrt{2})$; the configuration of altermagnetic Fermi surfaces is depicted below each of the panels. In all panels, we fix the electric potentials at the contacts as $e\bar{\phi}_{1,\lambda}/t_0 = -1$ and $e\bar{\phi}_{2,\lambda}/t_0 = 1$. The currents are normalized by $j_{0} = 2e\tau t_0 \, \mbox{max}{\left\{\left|\rho_{\lambda}^{(0)} \left(\bar{\phi}_{2,\lambda}-\bar{\phi}_{1,\lambda}\right) \right|\right\}}/L$, where the maximal value is taken with respect to the spin projections, and we fixed $\rho^{(0)}_{\uparrow}=\rho^{(0)}_{\downarrow}=\rho^{(0)}/2$. Red lines denote the position of the source and drain.
}
\label{fig:kinetic-current-am}
\end{figure*}

\textit{Transport in finite $d$-wave altermagnets.}
Let us evaluate electric and spin response in a $d$-wave altermagnet with $J(\mathbf{k}) = t_1(k_x^2-k_y^2) +2t_2 k_x k_y$. Then, Eq.~\eqref{eq:nonlocal-eq-fin} acquires a particularly simple form,
\begin{equation}
\label{eq:Ohmic-phi-lambda-eq}
t_{0} \Delta\bar{\phi}_{\lambda} + \lambda t_1 \left( \nabla_{x}^2 - \nabla_{y}^2 \right)  \bar{\phi}_{\lambda} +2 \lambda t_2 \nabla_{x} \nabla_{y}  \bar{\phi}_{\lambda} =0.
\end{equation}
This differential equation should be supplemented with boundary conditions. We fix $\bar{\phi}_{\lambda}$ at the contacts, corresponding to an electric or spin voltage bias~\footnote{A different configuration of the contacts where the source and drain are located at the same surface is considered in the Supplemental Material~\cite{SM}},
\begin{equation}
\label{eq:phi-lambda-gen-bc}
\begin{aligned}
\bar{\phi}_{\lambda} \left(x=0,y\right)&=\bar{\phi}_{1,\lambda}(y),\,\,|y|<w,\\
\bar{\phi}_{\lambda} \left(x=L_x,y\right)&=\bar{\phi}_{2,\lambda}(y),\,\,|y|<w,
\end{aligned}
\end{equation}
where $L_x$ is the length of the sample in the $x$-direction and $w$ is the width of the contacts located at $x=0$ and $x=L_x$ sides of the rectangular sample. The component of the electric and spin current normal to the surface vanishes everywhere outside the contacts. In our numerical calculations, we use the finite element method implemented in Wolfram \textit{Mathematica}~\cite{Wolfram}.

The electric and spin current densities for a few values of parameters $t_1$ and $t_2$ at $L_x=L_y=L$ are shown in Fig.~\ref{fig:kinetic-current-am} for the case of an applied electric voltage bias. The electric current streamlines are only weakly affected by the altermagnetic parameters, and hence we present only a single plot in Fig.~\ref{fig:kinetic-current-am}(a), which shows a typical distribution of the electric current for the Ohmic transport regime. The spin current, however, is nontrivial and depends strongly on the altermagnetic parameters $t_1$ and $t_2$, which quantify the magnitude of the spin splitting and the orientation of the altermagnetic nodal planes, see Figs.~\ref{fig:kinetic-current-am}(b)--\ref{fig:kinetic-current-am}(d). First of all, vortices of the spin current are observed at the sides of the contacts, independent of the orientation of the nodal planes. These vortices originate from the anisotropy of $T_{\lambda,ii}$. The details of the spin current flow are sensitive to the orientation of the altermagnetic nodal planes. In the case when at least one of the planes is perpendicular to the surface with the contacts (e.g., $t_1=0$ and $t_2\neq0$), the spin current forms a loop that begins and ends at the same contact, see Fig.~\ref{fig:kinetic-current-am}(c). This type of swirling current relies on the off-diagonal components of $T_{\lambda,ij}$. The spin current away from the contacts is directed perpendicular to the electric field, whose field lines coincide with those of the electric current in Fig.~\ref{fig:kinetic-current-am}(a), confirming the spin splitter effect~\cite{Naka-Seo-SpinCurrentGeneration-2019, Gonzalez-Hernandez-Zelezny:2021}. These papers, however, do not consider finite geometries and nonuniform electric fields.

In the case under consideration, the equations describing the distribution of the electrochemical potential \eqref{eq:nonlocal-eq-fin} and \eqref{eq:phi-lambda-gen-bc} are symmetric with respect to the interchange of the spin species. Therefore, similar to the swirling spin currents shown in Figs.~\ref{fig:kinetic-current-am}(b)--\ref{fig:kinetic-current-am}(d), one can obtain swirling {\sl electric} currents if the spin imbalance is created at the contacts. The corresponding currents have the same shape and magnitude as those in Fig.~\ref{fig:kinetic-current-am} with $j_{\rm el} \leftrightarrow j_{\rm sp}$. The spin imbalance can be generated via, e.g., the spin Hall effect~\cite{DYakonov-Perel-PossibilityOrientingElectron-1971, Hirsch-SpinHallEffect-1999, Hoffmann-SpinHallEffects-2013} or via spin-pumping \cite{tserkovnyak_rmp_04}.

\begin{figure*}[t]
\centering
\includegraphics[width=0.75\textwidth]{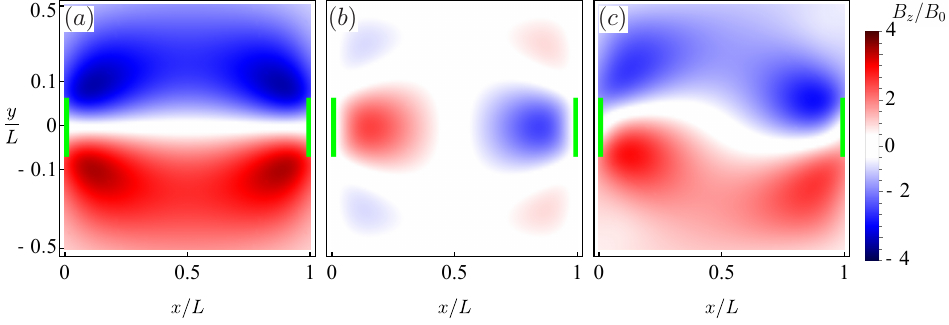}
\caption{
The spatial dependence of the induced magnetic field $B_z(\mathbf{r},z)$ for $z=0.1\,L$ at (a) $t_1=t_0/2$ and $t_2=0$, (b) $t_1=0$ and $t_2=t_0/2$, and (c) $t_1=t_2=t_0/(2\sqrt{2})$. In all panels, we fixed the potentials at the contacts as $e\bar{\phi}_{1,\lambda}/t_0 = -\lambda$ and $e\bar{\phi}_{2,\lambda}/t_0 = \lambda$, i.e., only the spin imbalance is applied, and we fixed $\rho^{(0)}_{\uparrow}=\rho^{(0)}_{\downarrow}=\rho^{(0)}/2$.
}
\label{fig:kinetic-B-am}
\end{figure*}

The possibility of generating swirling electric currents instead of the spin currents opens the door to the observation of the proposed effects via stray magnetic fields. A similar approach was successfully used in mapping out the electric current profile in the electron hydrodynamics~\cite{Ku-Walsworth:2019, Vool-Yacoby:2020-WTe2, Kumar-Ilani:2021, Jenkins-BleszynskiJayich-ImagingBreakdownOhmic-2022, Aharon-Steinberg2022}. We provide the distribution of the current-induced magnetic field in Fig.~\ref{fig:kinetic-B-am} for three configurations of the altermagnetic splitting that allows for the current vortices; see the SM~\cite{SM} for the definition of the field.

While we focused on the particular model of $d$-wave altermagnets, other models of altermagnets should have similar distributions of the electric and spin currents as long as they allow for off-diagonal or anisotropic diagonal components of the altermagnetic part of the conductivity tensor (see the SM~\cite{SM}). The only difference will be in the values of the coefficients $T_{\lambda, ij}$ entering the conductivity tensor \eqref{eq:Pi-0-ij}.

\textit{Keldysh approach.}
To show that the observed swirling currents are not an artifact of the low-energy model, we consider a fully quantum mechanical lattice model of a $d$-wave altermagnet and apply the Keldysh formalism~\cite{Keldysh-DiagramTechniqueNonequilibrium-1964, Nikolic-Souma-ImagingMesoscopicSpin-2006}, which was recently used to visualize the spin splitter effect in a homogeneous electric field~\cite{Hallberg-Linder-VisualizationSpinsplitterEffect-2025}.

A $d_{xy}$-altermagnet is modeled by the following tight-binding Hamiltonian:
\begin{equation}
    H_S = \sum_{i \sigma} \varepsilon_i c^\dagger_{i\sigma}c_{i \sigma} + \sum_{ij\sigma \sigma'} c_{i\sigma}^\dagger t_{ij}^{\sigma \sigma'}c_{j\sigma'},
\end{equation}
where $c_{i\sigma}$ ($c_{i\sigma}^{\dag}$) is the annihilation (creation) operator of electrons with the spin $\sigma$ at the site $i$. The onsite potential is $\varepsilon_i$ and the spin-dependent hopping terms
\begin{equation}
    \hat{t}_{ij} = \begin{cases}
    -t \mathbb I, \quad \left(j = i \pm e_x \; \mathrm{or} \; j = i \pm e_y\right) \\
    -t_m \sigma_z, \quad \left[ j = i \pm \left(e_x + e_y\right)\right] \\
    t_m \sigma_z, \quad \left[j = i \pm\left(e_x - e_y\right)\right],\end{cases}
\end{equation}
where $e_i$ is the unit vector in the $i$-th direction and $t_m$ corresponds to the strength of the spin-dependent diagonal hopping. The leads and connection between the leads and sample region are modelled as metals with $\varepsilon_i = 0$, and we only consider nearest neighbor hopping $t_{ij}^{\mathrm{leads}} = -t \mathbb I$ and $t_{ij}^{\mathrm{interface}} = -t \mathbb I$, respectively. The constrictions between the leads and the sample are modelled by a very high on-site potential $\varepsilon_{\mathrm{edge}} = 10^5\,t$ on the edges of the sample, except for a region of $10$ sites where the on-site potential is zero.

We define a bond spin-current operator representing the flow of spin $S_k$ from site $i$ to $j$ via
\begin{equation}
    J_{ij}^{S_k} \equiv \frac{1}{4i}\sum_{\alpha \beta}\left(c_{j\beta}^\dagger \{\sigma_k, \hat{t}_{ji} \}_{\beta \alpha} c_{i \alpha} - \mathrm{H.c.}\right),
\end{equation}
where the curly brackets denote the anticommutator. The steady-state non-equilibrium statistical average of this current is calculated in the Keldysh formalism (see the SM~\cite{SM}).

To include disorder in the altermagnet, we uniformly assign the fraction $n_I = 0.2$ of the sites with a local potential of the strength $\varepsilon_{\rm imp} = 10\,t_m$ simulating pointlike impurities. The disorder average is taken with respect to $100$ configurations of impurities.

\begin{figure}[t]
\centering
\includegraphics[width=0.4\textwidth]{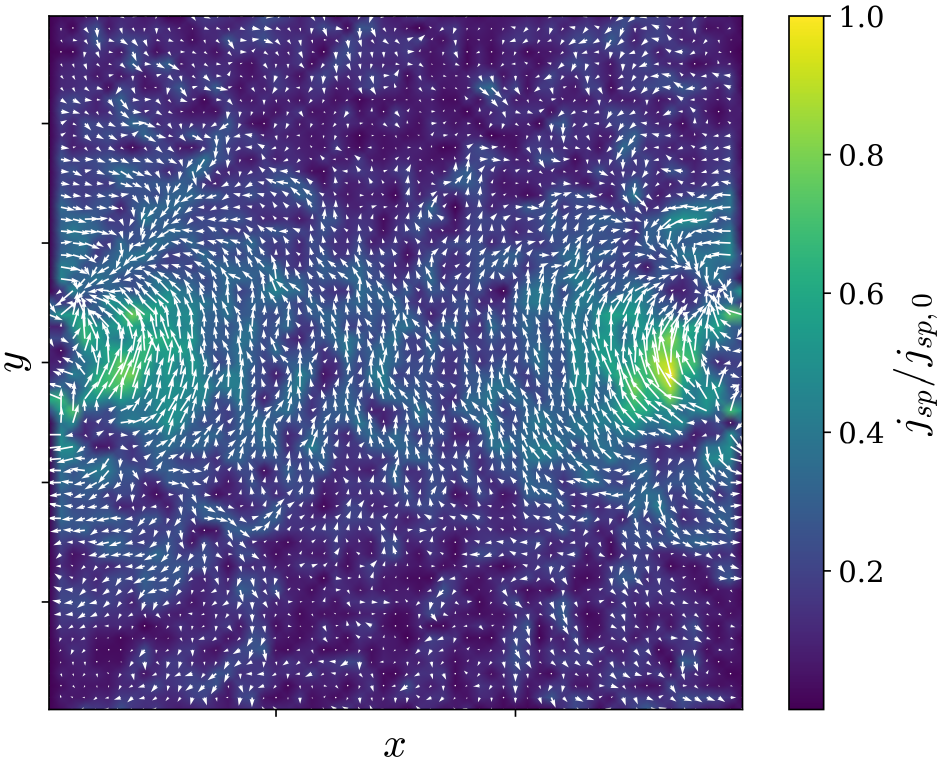}
\caption{
Spin current distribution obtained via the Keldysh formalism. The contacts of the width $w=10\,a$ are at $x=0,L_x$ surfaces. We use $60\times 60$ lattice with the following parameters: $t_m = 0.3\,t$, $\mu = -2.0\,t$, and the voltage bias $eV = 0.2\,t$.
}
\label{fig:Keldysh}
\end{figure}

The results of the lattice calculations in the disordered regime show a good qualitative agreement with the current distribution obtained in the kinetic approach; cf. Figs.~\ref{fig:kinetic-current-am}(c) and \ref{fig:Keldysh}. While the side vortices are hard to resolve due to the low magnitude of the current there, the central vortices look similar in both approaches. Vortical currents can also be observed in the ballistic transport regime, albeit the corresponding vorticity is not unique to altermagnets (see the SM~\cite{SM}).

\textit{Material estimates.}
The characteristic values of the key quantities, i.e., magnetic field and current, are
\begin{eqnarray}
\label{estimate-B0-phi}
B_0[n\mbox{T}]\! &=& \! \frac{73.9}{\sqrt{1 \! -(t_1^2+t_2^2)/t_0^2}} \frac{\mu \left[\mbox{meV}\right]\! \tau[\mbox{ps}] \delta \phi[\text{mV}]}{L[\mu\text{m}]},\\
\label{estimate-j0-phi}
j_0 \left[\frac{\mbox{A}}{\mbox{m}}\right]\! &=&  \! \frac{0.06}{\sqrt{1\! -(t_1^2+t_2^2)/t_0^2}} \frac{\mu\left[\mbox{meV}\right]\! \tau[\mbox{ps}] \delta \phi[\text{mV}]}{L[\mu\text{m}]},
\end{eqnarray}
where $\delta \phi =\bar{\phi}_{2}-\bar{\phi}_{1}$ is the applied voltage difference.

To estimate the observed effects, we use the material parameters of altermagnetic candidates such as RuO$_2$~\footnote{While the recent studies suggest that RuO$_2$ is nonmagnetic~\cite{Hiraishi-Hiroi-NonmagneticGroundState-2024, Kessler-Moser-Ru02:2024, Kiefer-Braden-CrystalStructureAbsence-2024}, it may support altermagnetism in thin films due to strain~\cite{Weber-Schneider-AllOpticalExcitation-2024, Jeong-Jalan-AltermagneticPolarMetallic-2024}.} and KRu$_4$O$_8$. We use $t_0a^{-2}=2.5~\mbox{eV}$, $t_1/t_0=t_2/t_0=0.2$, and $\mu=0.4~\mbox{eV}$ for RuO$_2$~\cite{Ahn-Kunes:2019, Smejkal-Jungwirth-GiantTunnelingMagnetoresistance-2022, Zhang-Neupert-FinitemomentumCooperPairing-2024} and $t_1/t_0=t_2/t_0=0.36$ and $\mu=0.05~\mbox{eV}$ for KRu$_4$O$_8$~\cite{Smejkal-Jungwirth-ConventionalFerromagnetismAntiferromagnetism-2022, Zhang-Neupert-FinitemomentumCooperPairing-2024}.

Assuming the width of ribbon $L=5~\mu\mbox{m}$, typical scattering time $\tau=1~\mbox{ps}$, and fixing the applied potential difference $\delta \phi = 1~\mbox{mV}$, we estimate $B_0 \approx 6~\mu\mbox{T}$ and the current density $j_0 \approx 24\, \mu \mbox{A/m}$ for RuO$_2$ as well as $B_0 \approx 1~\mu\mbox{T}$ and the current density $j_0 =3\, \mu \mbox{A/m}$ for KRu$_4$O$_8$. These values are small but are within reach of modern magnetic field imaging techniques such as nanowire magnetic force microscopy~\cite{Mattiat-Poggio-NanowireMagneticForce-2020, Marchiori-Poggio-NanoscaleMagneticField-2021}, the scanning superconducting quantum interference device (SQUID) magnetometry~\cite{Cui-Moler:2017, Marchiori-Poggio-NanoscaleMagneticField-2021}, and also at the edge of the sensitivity of the quantum spin magnetometry~\cite{Maze-Lukin-NanoscaleMagneticSensing-2008, Levine-Walsworth-PrinciplesTechniquesQuantum-2019, Marchiori-Poggio-NanoscaleMagneticField-2021, Rovny-Leon-NewOpportunitiesCondensed-2024}. In addition to the magnetic field, one can measure the local distribution of the electric potential, albeit the latter is less sensitive to the nontrivial distribution of currents~\cite{SM}.

\textit{Summary and conclusion.}
In this Letter, we showed that the altermagnetic spin splitting in finite-size samples with point contacts allows for swirling electric and spin currents even in disordered systems; i.e., in the Ohmic transport regime. The shape of the vortices encodes the relative orientation of the sample boundaries and the altermagnetic nodal planes, see Fig.~\ref{fig:kinetic-current-am}. Swirling electric (spin) currents can be generated if the spin imbalance (voltage) is applied to the contacts, as is supported by Eqs.~\eqref{eq:vorticity-eq-el} and \eqref{eq:vorticity-eq-sp}. The nontrivial distribution of the electric current is manifested in the induced magnetic field and is within reach of modern magnetometry, complementing previous transport measurements.

Comparing semiclassical calculations in a continuum model with the Keldysh approach on a lattice, we see qualitatively similar results supporting the model-independence of our findings; cf. Figs.~\ref{fig:kinetic-current-am}(c) and \ref{fig:Keldysh}. We emphasize that our results do not require strong electron-electron interaction (hydrodynamic transport regime) or ultrapure samples (ballistic transport regime), and take place in the Ohmic transport regime with a fully \emph{local} response to external stimuli. This broadens the selection of material platforms.

\begin{acknowledgments}
{\it Acknowledgments.}
A.H. acknowledges support from Project No.~0122U000887 of the Department of Physics and Astronomy of the NAS of Ukraine.
The work of E.V.G. was supported by the Program “Dynamics of particles and collective excitations in high-energy physics, astrophysics and quantum macrosystems” (Grant No. 0121U109612) of the Department of Physics and Astronomy of the NAS of Ukraine.
This work was supported by the Research Council of Norway through Grant No. 353894 and its Centres of Excellence funding scheme Grant No. 262633 “QuSpin.” Support from Sigma2 - The National Infrastructure for High Performance Computing and Data Storage in Norway, Project NN9577K, is acknowledged.
P.O.S. acknowledges useful communications with Alireza Qaiumzadeh and Xanthe Verbeek, and is grateful to the QuSpin center of excellence for warm hospitality.
We thank Sebasti\'{a}n Bergeret for reminding us about Ref.~\cite{Kokkeler-Bergeret-QuantumTransportTheory-2025}.
\end{acknowledgments}

\bibliography{library-short}

\end{document}

%\date{\today}

\title{Supplemental Material for \\
\textit{Electric and spin current vortices in altermagnets}}

\author{Arsen~Herasymchuk}
\email{arsengerasymchuk@gmail.com}
\affiliation{Bogolyubov Institute for Theoretical Physics, Kyiv, 03143, Ukraine}
%\affiliation{Department of Physics, Taras Shevchenko National Kyiv University, Kyiv, 01601, Ukraine}

\author{Karl Bergson Hallberg}
\affiliation{Center for Quantum Spintronics, Department of Physics, Norwegian University of Science and Technology, NO-7491 Trondheim, Norway}

\author{Erik Wegner Hodt}
\affiliation{Center for Quantum Spintronics, Department of Physics, Norwegian University of Science and Technology, NO-7491 Trondheim, Norway}

\author{Jacob Linder}
\affiliation{Center for Quantum Spintronics, Department of Physics, Norwegian University of Science and Technology, NO-7491 Trondheim, Norway}

\author{E. V. Gorbar}
\email{gorbar@knu.ua}
\affiliation{Department of Physics, Taras Shevchenko National Kyiv University, Kyiv, 01601, Ukraine}
\affiliation{Bogolyubov Institute for Theoretical Physics, Kyiv, 03143, Ukraine}

\author{Pavlo~Sukhachov}
\email{pavlo.sukhachov@missouri.edu}
\affiliation{Department of Physics and Astronomy, University of Missouri, Columbia, Missouri, 65211, USA}
\affiliation{MU Materials Science \& Engineering Institute,
University of Missouri, Columbia, Missouri, 65211, USA}

\maketitle

\tableofcontents

\section{Transport equations in altermagnets}
\label{sec:sm-transport}

In this section, we present the details of the transport equations in the semiclassical Boltzmann formalism.

\subsection{General equations}
\label{sec:sm-transport-gen}

A continuum model for a two-dimensional (2D) altermagnet reads
\begin{equation}
H= t_0 k^2 + \sigma_{z} J(k_x , k_y),
\end{equation}
where $k=\sqrt{k_x^2+k_y^2}$, $\sigma_i$ are the Pauli matrices in the spin space, and $J(k_x,k_y)$ is a function that encodes the $d$-, $g$-, or $i$- symmetry of the altermagnet. The dispersion relation for quasiparticles with the spin projection $\lambda$ is
\begin{equation}
\label{sm-transport-gen-eps}
\varepsilon_{\lambda}= t_0 k^2 + \lambda J(k_x , k_y).
\end{equation}

The transport equations are obtained as moments of the Boltzmann equation given in Eq.~(A1), multiplied by $e$ for the continuity equation and $e\mathbf{v}_{\mathbf{k},\lambda}$ for the equation for the current $\mathbf{j}_{\lambda}$. The resulting equations are presented in Eqs.~(A5) and (3), which, for convenience, we present here,
\begin{eqnarray}
\label{eq:transport-eq-rho-lambda}
&&\frac{\partial \rho_\lambda}{\partial t} + \left( \bm{\nabla} \cdot \mathbf{j}_\lambda \right) = 0,\\
\label{eq:transport-eq-current-lambda}
&&\frac{\partial \mathbf{j}_\lambda}{\partial t} + \nabla_{i} \tilde{\Pi}_{ij} \mathbf{e}_{j}+e E_i \Pi_{ij} \mathbf{e}_{j} = - \frac{\mathbf{j}_\lambda}{\tau},
\end{eqnarray}
where the charge and current densities for quasiparticles with the spin projection $\lambda$ are defined in Eqs.~(A3) and (A4), i.e.,
\begin{eqnarray}
\label{eq:transport-rho-def}
\rho_\lambda (t,\mathbf{r})&=& e\int\frac{d^2k}{(2\pi)^2} f_{\lambda}(t,\mathbf{r},\mathbf{k}),\\ 
\label{eq:transport-current-def}
\mathbf{j}_\lambda (t,\mathbf{r}) &=& e\int\frac{d^2k}{(2\pi)^2} \mathbf{v}_{\mathbf{k},\lambda} f_{\lambda} (t,\mathbf{r},\mathbf{k}),
\end{eqnarray}
and we used the relaxation time approximation for the collision integral with $\tau$ being the elastic scattering time off nonmagnetic disorder. The other shorthand notations are
\begin{eqnarray}
\label{eq:tPi-ij}
\tilde{\Pi}_{ij}(t,\mathbf{r}) &=& e\int\frac{d^2k}{(2\pi)^2} v_{\mathbf{k},\lambda,i} v_{\mathbf{k},\lambda,j} f_{\lambda} (t,\mathbf{r},\mathbf{k}),\\
\label{eq:Pi-ij}
\Pi_{ij}(t,\mathbf{r}) &=& -e\int\frac{d^2k}{(2\pi)^2} (\partial_{k_i} \partial_{k_j} \varepsilon_{\lambda}) f_{\lambda} (t,\mathbf{r},\mathbf{k}).
\end{eqnarray}

For weak deviations of the distribution function from its equilibrium value $f^{(0)}_{\lambda} (\mathbf{k}) =1/\left[e^{\left(\varepsilon_{\lambda} -\mu_{\lambda} \right)/T}+1\right]$ with $T$ being temperature and $\mu_{\lambda}$ being a spin-dependent chemical potential, we linearize the transport equations. Neglecting temperature deviations, we obtain
\begin{equation}
\begin{aligned}
\tilde{\Pi}_{ij}(t,\mathbf{r}) &= e\int\frac{d^2k}{(2\pi)^2} v_{\mathbf{k},\lambda,i} v_{\mathbf{k},\lambda,j} f_{\lambda} (t,\mathbf{r},\mathbf{k}) \approx e\int\frac{d^2k}{(2\pi)^2} v_{\mathbf{k},\lambda,i} v_{\mathbf{k},\lambda,j} \left[ f^{(0)}_{\lambda} (\mathbf{k})  -\frac{\partial f_{\lambda}}{\partial \varepsilon_{\lambda}} \delta \mu_{\lambda} (t,\mathbf{r})\right] \\
&=\tilde{\Pi}_{ij}^{(0)}+e\int\frac{d^2k}{(2\pi)^2} (\partial_{k_i} \partial_{k_j} \varepsilon_{\lambda}) f^{(0)}_{\lambda} (\mathbf{k}) \delta \mu_{\lambda}(t,\mathbf{r}) = \tilde{\Pi}_{ij}^{(0)}-\Pi_{ij}^{(0)} \delta \mu_{\lambda}(t,\mathbf{r}).
\end{aligned}
\end{equation}
The linearized equation for $\bm{j}_{\lambda}$ reads
\begin{equation}
\label{eq:transport-eq-current-lambda-lin}
\frac{\partial j_{\lambda,j}}{\partial t}  - e\Pi_{ij}^{(0)}\nabla_{i} \bar{\phi}_{\lambda}  = - \frac{j_{\lambda,j}}{\tau},
\end{equation}
where $\bar{\phi}_{\lambda} = \delta \mu_{\lambda}/e +\phi$ is the spin-resolved electrochemical potential and $\mathbf{E}=-\bm{\nabla} \phi$. 

Re-arranging the terms in Eq.~\eqref{eq:transport-eq-current-lambda-lin} and assuming the steady-state regime, we obtain
\begin{equation}
\label{eq:j-eq-3}
e\nabla_{j} \bar{\phi}_{\lambda}  = \left(\Pi^{(0)}\right)^{-1}_{ij}\frac{j_{\lambda,i}}{\tau}.
\end{equation}

We use the following relation valid for all symmetries of altermagnets:
\begin{equation}
\label{eq:Pi-0-ij}
\Pi_{ij}^{(0)}= -2 \rho^{(0)}_{\lambda} \left( t_0  \delta_{ij} + \lambda T_{\lambda, ij} \right),
\end{equation}
where the matrix $T_{\lambda, ij}$ is defined as follows
\begin{equation}
\label{eq:tlij-def}
T_{\lambda, ij} = \frac{e}{2 \rho^{(0)}_{\lambda} } \int\frac{d^2k}{(2\pi)^2} \left[ \partial_{k_i} \partial_{k_j} J(\mathbf{k}) \right] f_{\lambda}^{(0)}(\mathbf{k})
\end{equation}
and $\rho_{\lambda}^{(0)}$ is given in Eq.~\eqref{eq:transport-rho-def} with $f_{\lambda}(t, \mathbf{r}, \mathbf{k}) \to f_{\lambda}^{(0)}(\mathbf{k})$. 
As follows from Eqs.~\eqref{eq:Pi-0-ij} and \eqref{eq:tlij-def}, the off-diagonal and spin-dependent structure of $\Pi_{ij}^{(0)}$ in Eq.~\eqref{eq:Pi-0-ij} originates from the altermagnetic spin splitting and results in a nontrivial relation between the current $\mathbf{j}_{\lambda}$ and the gradient of the electrochemical potential $\bm{\nabla} \bar{\phi}_{\lambda}$. This agrees with the constitutive relations obtained in Ref.~\cite{Kokkeler-Bergeret-QuantumTransportTheory-2025} via a different method.

In the case where the potential is fixed at the contacts, we take the divergence of Eq.~\eqref{eq:transport-eq-current-lambda-lin} and rewrite it in terms of $\bar{\phi}_{\lambda}$ as
\begin{equation}
\label{eq:transport-eq-current-lambda-lin-phi}
t_0\nabla^2  \bar{\phi}_{\lambda} + \lambda \left( T_{\lambda, xx} \nabla_{x}^2 +T_{\lambda, yy} \nabla_{y}^2 +2T_{\lambda, xy} \nabla_{x} \nabla_{y}\right) \bar{\phi}_{\lambda} =0.
\end{equation}
For the fixed current boundary conditions, we introduce the scalar flow function $\psi_{\lambda}\left( \mathbf{r}\right)$ as $\mathbf{j}_{\lambda} \left(\mathbf{r}\right) =\left\{ -\nabla_{y} \psi_{\lambda} \left( \mathbf{r}\right), \nabla_{x} \psi_{\lambda} \left( \mathbf{r}\right) \right\}$ that satisfies the following equation:
\begin{equation}
\label{eq:psi-lambda-eq-general}
t_0\nabla^2 \psi_{\lambda}+ \lambda \left( T_{\lambda, xx} \nabla_{y}^2 +T_{\lambda, yy} \nabla_{x}^2\right)\psi_{\lambda}+ \lambda \left( T_{\lambda, xy} +T_{\lambda, yx} \right) \nabla_{x} \nabla_{y} \psi_{\lambda}=0.
\end{equation}

\subsection{Models of altermagnets}
\label{sec:sm-transport-models}

In this section, we calculate the tensor $\hat{T}_{\lambda}$ for a few other models of altermagnetic spin splitting.

\subsubsection{Low-energy models}
\label{sec:sm-transport-models-1}

We use the following low-energy models~\cite{Smejkal-Jungwirth-ConventionalFerromagnetismAntiferromagnetism-2022}:
\begin{eqnarray}
\label{sm-transport-models-d}
d-\mbox{wave}: &\quad& J(\mathbf{k}) = t_1(k_x^2 -k_y^2) +2t_2 k_x k_y,\\
\label{sm-transport-models-g}
g-\mbox{wave}: &\quad& J(\mathbf{k})=t_1 k_x k_y \left( k_x^2-k_y^2\right) -\frac{1}{4} t_2\left[ \left( k_x^2-k_y^2\right)^2 -4k_x^2k_y^2\right],\\
\label{sm-transport-models-i}
i-\mbox{wave}: &\quad& J(\mathbf{k})=t_1 k_xk_y\left(3k_x^2-k_y^2\right)\left(3k_y^2-k_x^2\right) +\frac{1}{2} t_2 \left( k_x^2-k_y^2\right) \left[ \left( k_x^2+k_y^2\right)^2 -16 k_x^2k_y^2\right].
\end{eqnarray}

The tensor $\hat{T}_{\lambda}$ acquires a particularly simple form for $d$-wave altermagnets,
\begin{equation}
\label{eq:sm-transport-dwave-T}
\hat{T}_{\lambda} = t_1 \sigma_z +t_2 \sigma_x.
\end{equation}

The structure of the altermagnetic splitting $J(\mathbf{k})$ for a $d$-wave altermagnet allows for a simple expression for the spin-resolved charge density $\rho_{\lambda}^{(0)}$,
\begin{equation}
\label{eq:transport-eq-rho0}
\rho_{\lambda}^{(0)} = -\frac{e T}{4 \pi \tilde{t}_0 } \text{Li}_{1} \left( -e^{\mu_{\lambda }/T}\right) \stackrel{T\to0}{=} -\frac{e \mu_{\lambda}}{4 \pi \tilde{t}_0}
\end{equation}
with $\tilde{t}_0 = \sqrt{t_0^2 -t_{1}^2 -t_{2}^2}$.

In the case of other models of altermagnets, the equations of motion have the same structure but with different values of the parameters $T_{\lambda, xx} = -T_{\lambda, yy}$ and $T_{\lambda, xy}$; the former equality is due to symmetry reasons~\cite{Smejkal-Jungwirth-ConventionalFerromagnetismAntiferromagnetism-2022}. Hence, we expect the current streamlines to be similar. 

For $g-$wave altermagnets, the parameters $T_{\lambda, xx}$ and $T_{\lambda, xy}$ read
\begin{eqnarray}
T_{\lambda, xx} &=& \frac{e}{2 \rho^{(0)}_{\lambda} } \int\frac{d^2k}{(2\pi)^2} \left[ 6 t_1 k_x k_y -3 t_2 \left(k_x^2-k_y^2\right) \right] f_{\lambda}^{(0)}=\frac{e}{2 \rho^{(0)}_{\lambda} } \int\frac{d^2k}{(2\pi)^2} 3 k^2 t \sin \left(2 \phi -4\phi_0 \right) f_{\lambda}^{(0)},\\
T_{\lambda, xy} &=& \frac{e}{2 \rho^{(0)}_{\lambda} } \int\frac{d^2k}{(2\pi)^2} \left[ 6 t_2 k_x k_y +3 t_1 \left(k_x^2-k_y^2\right) \right] f_{\lambda}^{(0)}=\frac{e}{2 \rho^{(0)}_{\lambda} } \int\frac{d^2k}{(2\pi)^2} 3 k^2 t \cos \left(2 \phi -4\phi_0 \right) f_{\lambda}^{(0)},
\end{eqnarray}
where we used the following parametrization: $t_1 = t \cos{(4\phi_0)}$ and $t_2 = t \sin{(4\phi_0)}$. For any value of $\phi_0$, the parameters $T_{\lambda, xx}$ and $T_{\lambda, yy}$ vanish.

For $i-$wave altermagnets, the parameters $T_{\lambda, xx}$ and $T_{\lambda, xy}$ can be similarly rewritten as
\begin{eqnarray}
T_{\lambda, xx} &=& -\frac{e}{2 \rho^{(0)}_{\lambda} } \int\frac{d^2k}{(2\pi)^2} 15 k^4 t \sin \left(4 \phi -6\phi_0 \right) f_{\lambda}^{(0)},\\
T_{\lambda, xy} &=& -\frac{e}{2 \rho^{(0)}_{\lambda} } \int\frac{d^2k}{(2\pi)^2} 15 k^4 t \cos \left(4 \phi -6\phi_0 \right) f_{\lambda}^{(0)},
\end{eqnarray}
where we used the following parametrization: $t_1 = t \cos{(6\phi_0)}$ and $t_2 = t \sin{(6\phi_0)}$. As for $g-$wave altermagnets, the parameters $T_{\lambda, xx}$ and $T_{\lambda, yy}$ are also vanishing.

Therefore, since $T_{\lambda, ij}$ is always vanishing for the low-energy models for $g$- and $i$-wave altermagnets given in Eqs.~\eqref{sm-transport-models-g} and \eqref{sm-transport-models-i}, respectively, there are no well-pronounced electric and spin current vortices in the center of the sample. On the other hand, weaker side vortices should still be allowed due to anisotropic diagonal components of $T_{\lambda, ij}$. This agrees with the fact that such models do not support the spin-splitter effect because, for corresponding spin Laue groups, the spin conductivity tensor vanishes~\cite{Belashchenko-GiantStraininducedSpin-2025}. The central current vortices in $g$- and $i$-wave altermagnets may appear, however, if the symmetry is reduced by applying strains~\cite{Belashchenko-GiantStraininducedSpin-2025, Karetta-Sinova-StrainControlledDwave-2025}.

\subsubsection{Role of sublattice degrees of freedom}
\label{sec:sm-transport-models-2}

In general, models of altermagnets should also include sublattice degrees of freedom~\cite{Roig2024, sumita2025arxiv}. To show that these degrees of freedom lead to the same structure of the spin-resolved conductivity tensor and, therefore, do not affect our transport results qualitatively, we use the following model of 2D altermagnet~\cite{sumita2025arxiv}:
\begin{equation}
\label{SM-sublattice-H}
%\begin{aligned}
H(\mathbf{k})=H_0(\mathbf{k})+H_{\text{mol}}(\mathbf{k}),
%\end{aligned}
\end{equation}
where 
\begin{equation}
%\begin{aligned}
H_0(\mathbf{k})=-4 t_1 \cos \left(\frac{k_x a}{2} \right) \cos \left(\frac{k_y a}{2} \right) \tau_{x} \otimes \sigma_{0}-2 t_2 \left[ \cos \left( k_x a \right) \cos \left( k_y a \right) \tau_{0}\otimes \sigma_{0}-\sin \left( k_x a \right)\sin \left( k_y a \right)  \tau_{z} \otimes \sigma_{0} \right] -\hat{\mu}
%\end{aligned}
\end{equation}
and $H_{\text{mol}}(\mathbf{k})$ is the Hamiltonian of the molecular field responsible for the altermagnetic order
\begin{equation}
%\begin{aligned}
H_{\text{mol}}(\mathbf{k})=-h \tau_{z} \otimes \sigma_{z}.  
%\end{aligned}
\end{equation}
Here, $\tau_{i}$ are the Pauli matrices for the sublattice space, $\sigma_{i}$ are the Pauli matrices for the spin space, $\hat{\mu}=\tau_{0}\otimes \text{diag}(\mu_{+},\mu_{-})$, $\mu_{\lambda}$ is the spin-resolved chemical potential for the spin projection $\lambda=\pm$. 

The dispersion relation of the Hamiltonian \eqref{SM-sublattice-H} reads as follows
\begin{equation}
%\begin{aligned}
\varepsilon_{\zeta,\lambda} =-\mu_\lambda -2 t_2 \cos (k_x a) \cos (k_y a)+\zeta \sqrt{ \left[ 2  t_2 \sin (k_x a) \sin (k_y a) - \lambda h\right]^2 +4 t_1^2 \left[1+ \cos \left(k_x a \right)\right]\left[1+ \cos \left(k_y a \right)\right]},
%\end{aligned}
\end{equation}
where $\zeta =\pm$. Since, in the main text, we focus on the case where the Fermi level is close to the band bottom, we linearize the Hamiltonian \eqref{SM-sublattice-H} around the $\Gamma$ point. Up to the second order in $ak$, the model Hamiltonian $H(\mathbf{k})$ and its dispersion relation are
%In this case, the model Hamiltonian $H$ and its dispersion relation in the low-energy limit up to $\mathbf{k}^2a^2$ are:
\begin{equation}
\label{eq:orbital-degree-Hamiltonian-low}
%\begin{aligned}
H(\mathbf{k}) \approx -4 t_1  \tau_{x} \otimes \sigma_{0}+\frac{t_1k^2a^2}{2}  \tau_{x} \otimes \sigma_{0}-2 t_2 \left[ \left( 1- \frac{k^2 a^2}{2} \right) \tau_{0}\otimes \sigma_{0}-k_x k_y a^2 \tau_{z} \otimes \sigma_{0} \right] -h \tau_{z} \otimes \sigma_{z} -\hat{\mu},
%\end{aligned}
\end{equation}
\begin{equation}
\label{eq:dispersion-orbital}
\begin{aligned}
\varepsilon_{\zeta,\lambda}(\mathbf{k}) &\approx -\tilde{\mu}_{\mathbf{k},\lambda} +\zeta \tilde{\varepsilon}_{\mathbf{k},\lambda} = -\mu_\lambda -2 t_2+t_2 k^2 a^2+\zeta \sqrt{ \left(2 t_2 k_x k_y a^2 - \lambda h \right)^2+t_1^2 \left(k^2a^2 /2-4\right)^2 } \\
&\approx \mu_\lambda -2 t_2 +\zeta \sqrt{16 t_1^2 +h^2}+\left( t_2 a^2 - \zeta \frac{2 t_1^2 a^2}{\sqrt{16 t_1^2 +h^2}} \right)k^2- \zeta \lambda \frac{2 t_2 h a^2}{\sqrt{16 t_1^2 +h^2}} k_x k_y,
\end{aligned}
\end{equation}
respectively.
By comparing the above dispersion relation with that in Eq.~(2) in the main text and Eq.~\eqref{sm-transport-gen-eps}, we see that the spin-structure of the energy bands is qualitatively the same in the models with and without sublattice degrees of freedom. 

Now, let us consider a spin-resolved conductivity tensor and find its structure using the Kubo approach. Assuming $T\to0$, the real part of the conductivity tensor is given by
\begin{equation}
%\begin{aligned}
\label{eq:sigma-Kubo}
\text{Re}\,\sigma^{(\lambda)}_{\alpha\beta}(\Omega) = \frac{1}{ \Omega }\int_{-\infty}^{+\infty}d\omega \left[ f(\omega)-f(\omega+\Omega) \right]  \int \frac{d\mathbf{k}}{(2\pi)^2} \text{Tr} \left[ \hat{j}_{\alpha} \hat{A}_{\lambda}(\omega;\mathbf{k}) \hat{j}_{\beta} \hat{A}_{\lambda}(\omega+\Omega ;\mathbf{k})\right],
%\end{aligned}
\end{equation} 
where $\hat{A}_{\lambda}(\omega;\mathbf{k})$ is the spectral function. For the Hamiltonian~(\ref{eq:orbital-degree-Hamiltonian-low}), we obtain
\begin{equation}
\begin{aligned}
\hat{A}_{\lambda}(\omega;\mathbf{k})&=\frac{1}{2}\left[ \delta \left( \omega - \tilde{\mu}_{\mathbf{k},\lambda} - \tilde{\varepsilon}_{\mathbf{k},\lambda } \right)+ \delta \left(  \omega -   \tilde{\mu}_{\mathbf{k},\lambda} + \tilde{\varepsilon}_{\mathbf{k},\lambda}  \right) \right] \tau_0\\
&+\frac{1}{2 \tilde{\varepsilon}_{\mathbf{k},\lambda}} \left[ \left( \frac{ t_1 k^2 a^2}{2}  -4 t_1 \right) \tau_{x}+\left( 2t_2 k_x k_y a^2   - \lambda h\right) \tau_{z}  \right] \left[ \delta \left( \omega -   \tilde{\mu}_{\mathbf{k},\lambda} - \tilde{\varepsilon}_{\mathbf{k},\lambda} \right)- \delta \left(  \omega -  \tilde{\mu}_{\mathbf{k},\lambda} + \tilde{\varepsilon}_{\mathbf{k},\lambda}  \right) \right]
\end{aligned}
\end{equation}
and $\hat{j}_{\alpha}$ is current operator
\begin{equation}
%\begin{aligned}
\hat{j}_{\alpha} = e t_1a^2 k_{\alpha} \tau_{x} +2 e t_2 a^2 \left( k_{\alpha} \tau_{0} +k_{\bar{\alpha}}  \tau_{z} \right) %,\,\,\alpha=(x,y),\,\,\bar{\alpha}=(y,x)
%\end{aligned}
\end{equation}
with $\alpha=\{x,y\}$ and $\bar{\alpha}=\{y,x\}$. Note that since the Hamiltonian \eqref{eq:orbital-degree-Hamiltonian-low} is block-diagonal in the spin space, one can introduce the spin-projection index $\lambda$ in Eq.~\eqref{eq:sigma-Kubo}.  

Therefore, in dc limit, we obtain
\begin{equation}
\label{eq:conductivity-orbital}
%\begin{aligned}
\text{Re}\,\sigma^{(\lambda)}_{\alpha\beta}(\Omega) =  \lim_{\Omega \rightarrow 0} \int \frac{d\mathbf{k}}{(2\pi)^2}  \text{Tr} \left[ \hat{j}_{\alpha} \hat{A}_{\lambda}(0;\mathbf{k}) \hat{j}_{\beta} \hat{A}_{\lambda}(\Omega ;\mathbf{k})\right],
%\end{aligned}
\end{equation} 
where
\begin{equation}
\begin{aligned}
\text{Tr} \left[ \hat{j}_{\alpha} \hat{A}_{\lambda}(0;\mathbf{k}) \hat{j}_{\beta} \hat{A}_{\lambda}(\Omega ;\mathbf{k})\right]&=\frac{e^2 a^4 \delta(\Omega) }{2 \tilde{\varepsilon}_{\mathbf{k},\lambda}^2 } \left[ \delta \left(  - \tilde{\mu}_{\mathbf{k},\lambda} - \tilde{\varepsilon}_{\mathbf{k},\lambda } \right) +\delta \left(  - \tilde{\mu}_{\mathbf{k},\lambda} + \tilde{\varepsilon}_{\mathbf{k},\lambda } \right) \right] \\
&\times \Big\{ \left[ \left(4 t_2^2+t_1^2\right) k_{\alpha}k_{\beta} \tilde{\varepsilon}_{\mathbf{k},\lambda}^2  +4 t_2^2 k_{\bar{\alpha}}k_{\bar{\beta}} \tilde{\varepsilon}_{\mathbf{k},\lambda}^2  \right] \\
&+t_1^2 \left(k^2a^2-8\right)^2   \left[ \left(4 t_2^2-t_1^2\right) k_{\alpha}  k_{\beta}    - 4 t_2^2 k_{\bar{\alpha}} k_{\bar{\beta}} \right]   \\
&+ \left[ \left(4 t_2^2 - t_1^2 \right)  k_{\alpha} k_{\beta} + 4t_2^2 k_{\bar{\alpha}}  k_{\bar{\beta}} \right]  \left(2 t_2 k_x k_y a^2 - \lambda h \right)^2 \\
&  +2  t_1^2 t_2  \left(  k_{\alpha}k_{\bar{\beta}} +k_{\bar{\alpha}}k_{\beta} \right)  \left(k^2a^2-8\right) \left(2 t_2 k_x k_y a^2 - \lambda h  \right) \Big\}\\
&+ \frac{e^2 a^4 \delta(\Omega) }{2 \tilde{\varepsilon}_{\mathbf{k},\lambda}^2 } \left[ \delta \left(  - \tilde{\mu}_{\mathbf{k},\lambda} - \tilde{\varepsilon}_{\mathbf{k},\lambda } \right) -\delta \left(  - \tilde{\mu}_{\mathbf{k},\lambda} + \tilde{\varepsilon}_{\mathbf{k},\lambda } \right) \right]  \\
&\times \left[ 8 t_2^2 \left(  k_{\alpha}k_{\bar{\beta}} +k_{\bar{\alpha}}k_{\beta} \right) \left(2 t_2 k_x k_y a^2 -\lambda h \right) \tilde{\varepsilon}_{\mathbf{k},\lambda} + 4 t_1^2 t_2 k_{\alpha}k_{\beta} \tilde{\varepsilon}_{\mathbf{k},\lambda} \left(k^2a^2 -8\right) \right].
\end{aligned}
\end{equation} 
Due to symmetry reasons, the off-diagonal $\alpha \neq \beta$ components of the conductivity tensor proportional to $\lambda$ can emerge only at $h\neq0$.  %Therefore, since the appearance of the current vortices relies on the structure of the spin-resolved conductivity tensor rather than the values of its components, the qualitative results presented in the main text remain unchanged.

\begin{figure}[h]
\centering
\includegraphics[width=.45\textwidth]{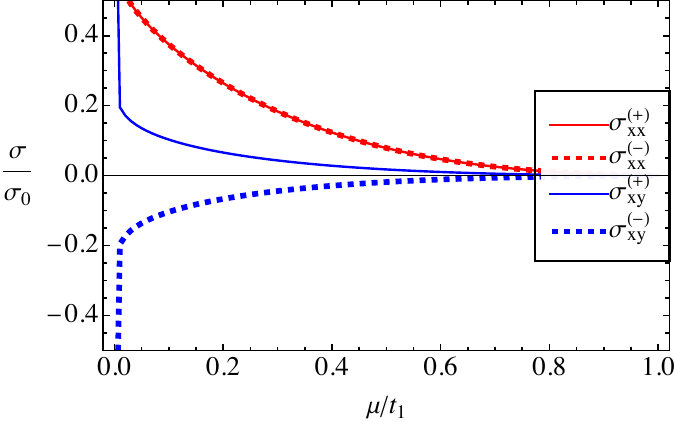}
\caption{
The dependence of the spin-resolved dc conductivity tensor $\sigma_{xx}^{(\lambda)}$ and $\sigma_{xy}^{(\lambda)}$ for the model \eqref{eq:orbital-degree-Hamiltonian-low} on $\mu/t_1$. We fix $t_2 = 0.2 \,t_1$ and $h = 0.4 \, t_1$, replaced $\delta(\Omega) \rightarrow \tau/ \pi$ with $\tau$ being the relaxation time, and defined $\sigma_0=e^2\mu \tau/\hbar^2$ (we restored $\hbar$ in this expression). Solid and dashed red lines correspond to $\sigma_{xx}^{(+)}$ and $\sigma_{xx}^{(-)}$, respectively. Solid and dashed blue lines correspond to $\sigma_{xy}^{(+)}$ and $\sigma_{xy}^{(-)}$, respectively.
Other components of the conductivity tensor satisfy $\sigma_{xx}^{(\lambda)}=\sigma_{yy}^{(\lambda)}$ and $\sigma_{xy}^{(\lambda)}=\sigma_{yx}^{(\lambda)}$.
}
\label{fig:conductivity-orb}
\end{figure}

The dependence of the spin-resolved dc conductivity tensor components on $\mu/t_1$ for $t_2=0.2\, t_1$, $h = 0.4\, t_1$ for both spin projections $\lambda=\pm$ is presented in Fig.~\ref{fig:conductivity-orb}. Note that $\sigma_{xx}=\sigma_{yy}$ and $\sigma_{xy}=\sigma_{yx}$, as expected for a $d$-wave altermagnet. These calculations reconfirm that the spin-resolved conductivity tensor in a model with sublattice degrees of freedom has the same structure as for the effective model considered in the main text. Therefore, since the appearance of the current vortices relies on the structure of the spin-resolved conductivity tensor rather than the values of its components, the qualitative results presented in the main text are also applicable to models with a sublattice degree of freedom.

\section{Fixed current at the contacts}
\label{sec:analyt}

In this section, we discuss the model setup that allows for analytical solutions. We consider $d$-wave altermagnet with the ribbon geometry, where $x\in \left(-\infty, \infty\right)$ and $y\in \left[0,L\right]$. The contacts with fixed current are applied to the opposite sides of the ribbon, i.e., we fix
\begin{equation}
\label{eq:analyt-j-bc}
j_{y,\lambda}(x,y=0) = \frac{j_{1,\lambda}(x)}{2}, \quad 
j_{y,\lambda}(x,y=L) = \frac{j_{2,\lambda}(x)}{2}.
\end{equation}
Another type of geometry that allows for analytical results is the vicinity geometry, where the contacts are located on the same side of the sample, i.e., with $j_{y,\lambda}(x,y=0)=0$ and $j_{y,\lambda}(x,y=L)=j_{2,\lambda}(x)/2$.

To solve Eq.~\eqref{eq:psi-lambda-eq-general}, we perform the Fourier transform with respect to $x$,
\begin{equation}
\label{eq:euler-psi-k-def}
\psi_{\lambda} \left(\mathbf{r} \right) =\frac{1}{2\pi} \int_{-\infty}^{+\infty} dk \, e^{ikx} \psi_{\lambda, k}(y)
\end{equation}
and seek the solution for $\psi_{\lambda, k}(y)$ as
\begin{equation}
\label{eq:psi-lambda}
\psi_{\lambda, k}(y)= \sum_{i=1,2} A_{i} e^{\varkappa_{i} y},
\end{equation}
where 
\begin{equation}
\label{eq:analyt-varkappa}
\varkappa_{1} =k \frac{\tilde{T}_0-i \lambda T_{xy}}{t_0-\lambda T_{xx}} \quad \mbox{and} \quad \varkappa_{2} =k \frac{-\tilde{T}_0-i \lambda T_{xy}}{t_0-\lambda T_{xx}}.
\end{equation}
Here, $\tilde{T}_0 = \sqrt{t_0^2 -T_{xx}^2 -T_{xy}^2}$ and we assumed that there is no spin density in the equilibrium, $\mu_{\lambda}=\mu$, hence $\hat{T}_{\lambda} = \hat{T}$.
Therefore, for the boundary conditions in Eq.~\eqref{eq:analyt-j-bc}, the stream function is
\begin{equation}
\label{eq:psi-lambda-fin}
\psi_{\lambda}(\mathbf{r})=\frac{1}{4 \pi i} \int_{-\infty}^{+\infty}dk \,  \frac{e^{ik\left(x-\frac{ \lambda T_{xy} }{t_0-\lambda T_{xx}} y \right)} }{k \sinh \left( \frac{\tilde{T}_0 k L}{t_0-\lambda T_{xx}} \right) }  \left[ j_{2,\lambda}(k) e^{ikL \frac{\lambda T_{xy} }{t_0-\lambda T_{xx}} }  \sinh \left( \frac{\tilde{T}_0 k y}{t_0-\lambda T_{xx}} \right) +  j_{1,\lambda}(k) \sinh \left( \frac{\tilde{T}_0 k (L-y) }{t_0-\lambda T_{xx}}  \right)\right].
\end{equation}

Using Eq.~\eqref{eq:psi-lambda-fin} in Eq.~(\ref{eq:j-eq-3}), we obtain the following spin-resolved electrochemical potential: 
\begin{equation}
\label{eq:bmu-lambda-fin}
\bar{\phi}_{\lambda}
=- \frac{1}{4 \pi e \tau \tilde{T}_0 \rho^{(0)} } \int_{-\infty}^{+\infty}dk \,  \frac{e^{ik\left(x-\frac{ \lambda T_{xy} }{t_0-\lambda T_{xx}} y \right)} }{k \sinh \left( \frac{\tilde{T}_0 k L}{t_0-\lambda T_{xx}} \right) } \left[ j_{2,\lambda}(k) e^{ikL \frac{\lambda T_{xy} }{t_0-\lambda T_{xx}} }  \cosh \left( \frac{\tilde{T}_0 k y}{t_0-\lambda T_{xx}} \right) %\right. \\&\left.
-  j_{1,\lambda}(k) \cosh \left( \frac{\tilde{T}_0 k (L-y) }{t_0-\lambda T_{xx}}  \right) \right].
\end{equation}

In what follows, we focus on the point-like contacts where $j_{1,\lambda}(x) =j_{2,\lambda}(x) =I \delta{(x)}$ for the contacts on the opposing sides and $j_{1,\lambda}(x)=0$ and $j_{2,\lambda}(x)=-I\delta\left(x-x_0\right)+I\delta\left(x+x_0\right)$ for the vicinity geometry. Here, $I$ is the total current supplied via the contacts.

\begin{figure}[h]
\centering
\includegraphics[width=.6\textwidth]{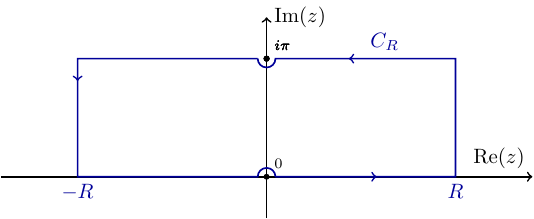}
\caption{Integration contour $C_{R}$ used in Eq.~\eqref{eq:analyt-JCR}.}
\label{fig:contour}
\end{figure}

In our analytical calculations, we use the following integral:
\begin{equation}
\label{eq:analyt-JCR}
J_{C_R}(a,b) =\oint_{C_{R}} dz\, \frac{e^{(a+ib)z}}{\sinh{z}},
\end{equation}
where $|a| < 1$ and the contour $C_{R}$ is shown in Fig.~\ref{fig:contour}.

In the limit $R \rightarrow +\infty$, we obtain 
\begin{equation}
\label{eq:cbc-current-gen}
J_{C_R}(a,b)=\int_{-\infty}^{+\infty} \,dx \frac{e^{(a+ib)x}}{\sinh{x}}= \pi i\frac{1- e^{i \pi a}e^{- \pi b}}{1+e^{i \pi a}e^{- \pi b}}.
\end{equation}

By using the relation \eqref{eq:cbc-current-gen} in Eq.~\eqref{eq:psi-lambda-fin},
we derive the following analytical expression for $\mathbf{j}_{\lambda} (\mathbf{r})$ in the geometry with opposing contacts:
\begin{equation}
\label{eq:cbc-current-x}
\begin{aligned}
j_{\lambda,x}(\mathbf{r})&=\frac{\lambda T_{xy}}{t_0-\lambda T_{xx}}j_{\lambda,y}(\mathbf{r})-\frac{I}{4L} \frac{\sinh\left( \frac{\pi }{\tilde{T}_0} \frac{\left( t_0 -\lambda T_{xx} \right)x -\lambda T_{xy} \left(y-L\right)}{ L }\right)}{\cosh\left( \frac{\pi }{\tilde{T}_0} \frac{\left( t_0 -\lambda T_{xx} \right)x -\lambda T_{xy} \left(y-L\right)}{ L }\right)+ \cos \left( \frac{\pi y}{L} \right)} \\
&+ \frac{I}{4L} \frac{\sinh\left( \frac{\pi }{\tilde{T}_0}\frac{\left( t_0 - \lambda T_{xx} \right)x - \lambda T_{xy} y}{ L }\right)}{\cosh\left( \frac{\pi }{\tilde{T}_0}\frac{\left( t_0 - \lambda T_{xx} \right)x - \lambda T_{xy} y}{ L }\right)- \cos \left( \frac{\pi y}{L} \right)},\\
\end{aligned}
\end{equation}
\begin{equation}
\label{eq:cbc-current-y}
j_{\lambda,y}(\mathbf{r}) = \frac{I}{4\tilde{T}_0L} \frac{\left(t_0 - \lambda T_{xx}\right) \sin \left( \frac{\pi y}{L} \right)}{\cosh\left( \frac{\pi }{\tilde{T}_0} \frac{\left( t_0 -\lambda T_{xx} \right)x -\lambda T_{xy} \left(y-L\right)}{ L }\right)+ \cos \left( \frac{\pi y}{L} \right)}+ \frac{I}{4\tilde{T}_0L} \frac{\left(t_0 - \lambda T_{xx}\right) \sin \left( \frac{\pi y}{L} \right)}{\cosh\left( \frac{\pi }{\tilde{T}_0}\frac{\left( t_0 - \lambda T_{xx} \right)x - \lambda T_{xy} y}{ L }\right)- \cos \left( \frac{\pi y}{L} \right)}.
\end{equation}

The expressions for $\psi_{\lambda} \left( \mathbf{r} \right)$ and $\bar{\phi}_{\lambda}\left( \mathbf{r} \right)$ are obtained from the results for $\mathbf{j}_{\lambda} (\mathbf{r})$ given in Eqs.~\eqref{eq:cbc-current-x} and \eqref{eq:cbc-current-y}, and read as
\begin{equation}
\label{eq:analyt-psi-sol-1}
\begin{aligned}
\psi_{\lambda}(\mathbf{r})&= \frac{I}{2\pi} \text{Arctan}\,\left(\tan \left( \frac{\pi y}{2 L} \right) \tanh\left( \frac{\pi \left( t_0 -\lambda T_{xx} \right)x - \pi \lambda T_{xy} (y-L)}{ 2 \tilde{T}_0 L }\right) \right)\\
&- \frac{I}{2\pi} \text{Arctan}\, \left(\tan \left( \frac{\pi y}{2  L} \right) \coth\left( \frac{\pi \left( t_0 -\lambda T_{xx} \right)x - \pi \lambda T_{xy} y}{ 2 \tilde{T}_0 L }\right) \right),
\end{aligned}
\end{equation}
\begin{equation}
\label{eq:analyt-phi-sol-1}
\bar{\phi}_{\lambda}\left( \mathbf{r} \right) =- \frac{I}{4 \pi e \tau \tilde{T}_0 \rho^{(0)} }  \log\left[ \frac{\cosh \left(\frac{\pi \left(t_0 -\lambda T_{xx}  \right) x - \pi \lambda t_{2} y}{\tilde{T}_0L }\right)-\cos \left(\frac{\pi  y}{L}\right)}{\cosh \left(\frac{\pi  \left(t_0-\lambda T_{xx}\right) x-\pi \lambda T_{xy} (y-L)}{\tilde{T}_0 L}\right)+\cos \left(\frac{\pi  y}{L}\right)} \right],
\end{equation}
where $\text{Arctan}(x)$ is the multivalued version of the inverse trigonometric function $\arctan(x)$ so $\psi_{\lambda}(\mathbf{r})$ is continuous function.

In the vicinity geometry, $j_{1,\lambda}(x)=0$ and $j_{2,\lambda}(x)=-I\delta\left(x-x_0\right)+I\delta\left(x+x_0\right)$. The current $\mathbf{j}_{\lambda} (\mathbf{r})$ is given by
\begin{equation}
\begin{aligned}
j_{\lambda,x}(\mathbf{r})&=\frac{\lambda T_{xy}}{t_0-\lambda T_{xx}}j_{\lambda,y}(\mathbf{r})-\frac{I}{4L} \frac{\sinh\left( \frac{\pi }{\tilde{T}_0} \frac{\left( t_0 -\lambda T_{xx} \right)(x+x_0) -\lambda T_{xy} \left(y-L\right)}{ L }\right)}{\cosh\left( \frac{\pi }{\tilde{T}_0} \frac{\left( t_0 -\lambda T_{xx} \right)(x+x_0) -\lambda T_{xy} \left(y-L\right)}{ L }\right)+ \cos \left( \frac{\pi y}{L} \right)}\\
&+\frac{I}{4L} \frac{\sinh\left( \frac{\pi }{\tilde{T}_0} \frac{\left( t_0 -\lambda T_{xx} \right)(x-x_0) -\lambda T_{xy} \left(y-L\right)}{ L }\right)}{\cosh\left( \frac{\pi }{\tilde{T}_0} \frac{\left( t_0 -\lambda T_{xx} \right)(x-x_0) -\lambda T_{xy} \left(y-L\right)}{ L }\right)+ \cos \left( \frac{\pi y}{L} \right)},\\
\end{aligned}
\end{equation}
\begin{equation}
\begin{aligned}
j_{\lambda,y}(\mathbf{r})&= \frac{I}{4\tilde{T}_0L} \frac{\left(t_0 - \lambda T_{xx}\right) \sin \left( \frac{\pi y}{L} \right)}{\cosh\left( \frac{\pi }{\tilde{T}_0} \frac{\left( t_0 - \lambda T_{xx} \right)(x+x_0) -\lambda T_{xy} \left(y-L\right)}{ L }\right)+ \cos \left( \frac{\pi y}{L} \right)}\\
&-\frac{I}{4\tilde{T}_0L} \frac{\left(t_0 - \lambda T_{xx}\right) \sin \left( \frac{\pi y}{L} \right)}{\cosh\left( \frac{\pi }{\tilde{T}_0} \frac{\left( t_0 -\lambda T_{xx} \right)(x-x_0) -\lambda T_{xy} \left(y-L\right)}{ L }\right)+ \cos \left( \frac{\pi y}{L} \right)}.\\
\end{aligned}
\end{equation}

The functions $\psi_{\lambda} \left( \mathbf{r} \right)$ and $\bar{\phi}_{\lambda}\left( \mathbf{r} \right)$ are given by
\begin{equation}
\label{eq:analyt-psi-sol-2}
\begin{aligned}
\psi_{\lambda}(\mathbf{r})&= \frac{I}{2\pi} \text{Arctan}\, \left(\tan \left( \frac{\pi y}{2 L} \right) \tanh\left( \frac{\pi \left( t_0 -\lambda T_{xx} \right)(x+x_0) - \pi \lambda T_{xy} (y-L)}{ 2  L }\right) \right)\\
&- \frac{I}{2\pi} \text{Arctan}\,\left(\tan \left( \frac{\pi y}{2 L} \right) \tanh\left( \frac{\pi \left( t_0 -\lambda T_{xx} \right)(x-x_0) - \pi \lambda T_{xy} (y-L)}{ 2  L }\right) \right),
\end{aligned}
\end{equation}
\begin{equation}
\label{eq:analyt-phi-sol-2} 
\bar{\phi}_{\lambda}\left( \mathbf{r} \right) =- \frac{I}{4 \pi e \tau \tilde{T}_0 \rho^{(0)} }  \log\left[ \frac{\cosh \left(\frac{\pi  \left(t_0-\lambda T_{xx}\right) (x-x_0)-\pi \lambda T_{xy} (y-L)}{\tilde{T}_0 L}\right)+\cos \left(\frac{\pi  y}{L}\right)}{\cosh \left(\frac{\pi  \left(t_0-\lambda T_{xx}\right) (x+x_0)-\pi \lambda T_{xy} (y-L)}{\tilde{T}_0 L}\right)+\cos \left(\frac{\pi  y}{L}\right)} \right].
\end{equation}

\begin{figure}[h]
\centering
\includegraphics[width=.99\textwidth]{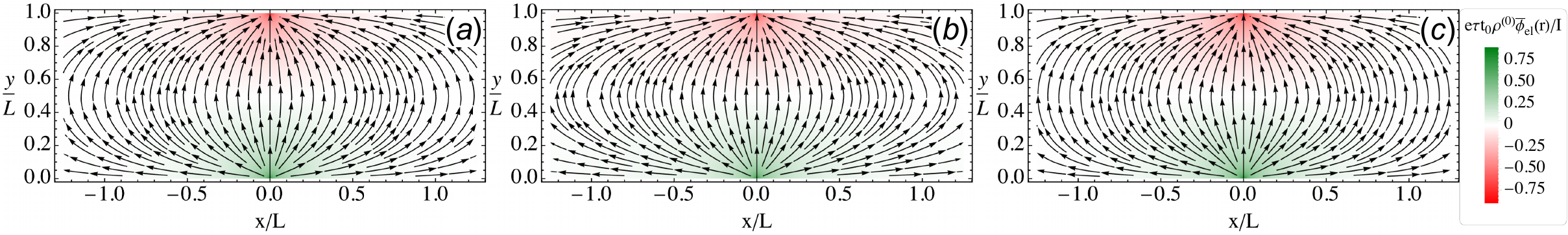}
\caption{
The electric current streamlines ($\psi_{\text{el}}(\mathbf{r}) = \text{const}$) for: (a) $t_{1,2}=0$; (b) $t_1=t_0/2 $, $t_2=0$; (c) $t_1=0$, $t_2=t_0/2 $. The electrochemical potential $e \tau \tilde{t}_0 \rho^{(0)} \bar{\phi}_{\text{el}} \left(\mathbf{r}\right) / I$ is shown in color. The green and red colors correspond to $\bar{\phi}_{\text{el}}(\mathbf{r})>0$ and $\bar{\phi}_{\text{el}}(\mathbf{r})<0$, respectively.
}
\label{fig:streamlines5}
\end{figure}

Let us visualize the obtained expression. For the sake of definiteness, we consider $d$-wave altermagnets with $T_{xx}=t_1$, $T_{xy}=t_2$, and $\tilde{T}_0 = \tilde{t}_0 =  \sqrt{t_0^2-t_{1}^2 - t_{2}^2}$. In the geometry where the source and drain are located symmetrically on opposite sides of the sample, the electric current streamlines with $\psi_{\text{el}}(\mathbf{r})=\sum_{\lambda} \psi_{\lambda}(\mathbf{r})$ and the distribution of the electrochemical potential $\bar{\phi}_{\text{el}}(\mathbf{r}) =\sum_{\lambda} \bar{\phi}_{\lambda}(\mathbf{r})$ are shown in Fig.~\ref{fig:streamlines5} at $\rho^{(0)}_{+}=\rho^{(0)}_{-}=\rho^{(0)}/2$. There are no well-pronounced changes due to altermagnetic spin splitting.

\begin{figure}[h]
\centering
\includegraphics[width=.99\textwidth]{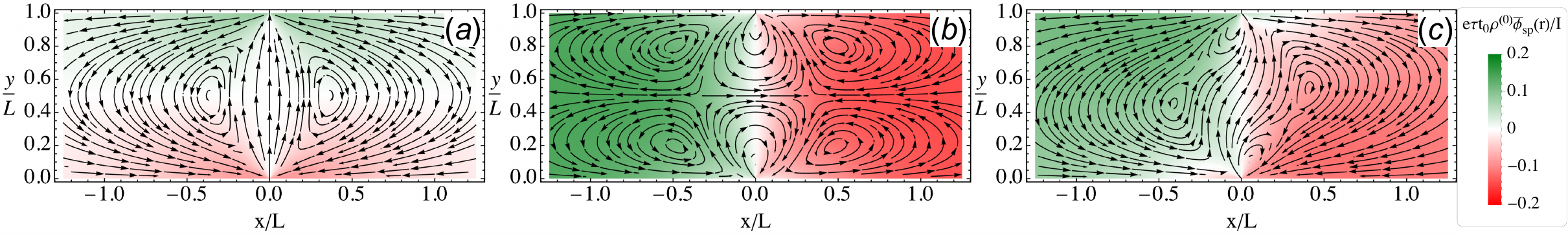}
\caption{
The spin current streamlines ($\psi_{\text{sp}}(\mathbf{r}) = \text{const}$) for: (a) $t_1=t_0/2 $, $t_2=0$; (b) $t_1=0 $, $t_2=t_0/2$; (c) $t_1=t_2=t_0/(2\sqrt{2})$. The spin electrochemical potential $e \tau \tilde{t}_0 \rho^{(0)} \bar{\phi}_{\text{sp}} \left(\mathbf{r}\right) / I$ is shown in color. The green and red colors correspond to $\bar{\phi}_{\text{sp}}(\mathbf{r})>0$ and $\bar{\phi}_{\text{sp}}(\mathbf{r})<0$, respectively.
}
\label{fig:streamlines6}
\end{figure}

The spin current streamlines with $\psi_{\text{sp}}(\mathbf{r})=\sum_{\lambda} \lambda \psi_{\lambda}(\mathbf{r})$ and the distribution of the spin electrochemical potential $\bar{\phi}_{\text{sp}}(\mathbf{r}) =\sum_{\lambda} \lambda \bar{\phi}_{\lambda}(\mathbf{r})$ are shown in Fig.~\ref{fig:streamlines6}  at $\rho^{(0)}_{+}=\rho^{(0)}_{-}=\rho^{(0)}/2$. The nonzero values of $t_{1,2}$ lead to nontrivial flow configurations of the spin current even without any spin current injection through the contacts. The obtained results, namely the presence of the vortices, are similar to the case with the fixed voltage considered in the main text. The major difference is in the structure of the side vortices, which are affected by the finite size in the $x$ direction in the setup used in the main text.

\begin{figure}[h]
\centering
\includegraphics[width=.99\textwidth]{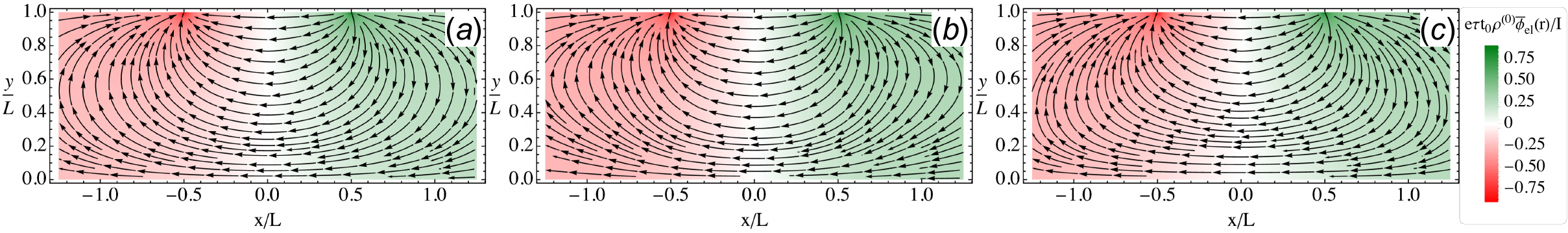}
\caption{
The electric current streamlines ($\psi_{\text{el}}(\mathbf{r}) = \text{const}$) at $x_0=L/2$ for: (a) $t_{1,2}=0$; (b) $t_1=t_0/2 $, $t_2=0$; (c) $t_1=0$, $t_2=t_0/2 $. The electrochemical potential $e \tau \tilde{t}_0 \rho^{(0)} \bar{\phi}_{\text{el}} \left(\mathbf{r}\right) / I$ is shown in color. The green and red colors correspond to $\bar{\phi}_{\text{el}}(\mathbf{r})>0$ and $\bar{\phi}_{\text{el}}(\mathbf{r})<0$, respectively.}
\label{fig:streamlines7}
\end{figure}

In the case of the vicinity geometry, i.e., when the source and drain are located on the same side of the channel, the electric current streamlines are shown in Fig.~\ref{fig:streamlines7}. As in Fig.~\ref{fig:streamlines5}, there are no well-pronounced changes in the electric current distribution in altermagnets.

\begin{figure}[h]
\centering
\includegraphics[width=.99\textwidth]{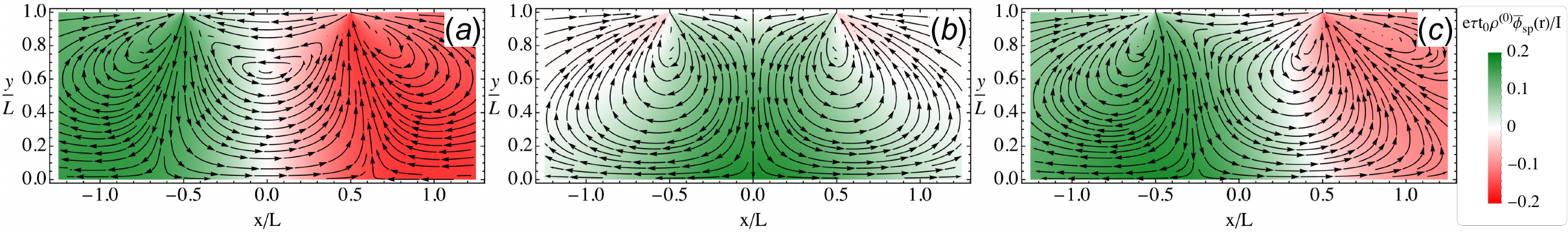}
\caption{
The spin current streamlines ($\psi_{\text{sp}}(\mathbf{r}) = \text{const}$) at $x_0=L/2$ for: (a) $t_1=t_0/2 $, $t_2=0$; (b) $t_1=0$, $t_2=t_0/2 $; (c) $t_{1}=t_2=t_0/(2\sqrt{2})$. The spin electrochemical potential $e \tau \tilde{t}_0 \rho^{(0)} \bar{\phi}_{\text{sp}} \left(\mathbf{r}\right) / I$ is shown in color. The green and red colors correspond to $\bar{\phi}_{\text{sp}}(\mathbf{r})>0$ and $\bar{\phi}_{\text{sp}}(\mathbf{r})<0$, respectively.}
\label{fig:streamlines8}
\end{figure}

The spin current streamlines with $\psi_{\text{sp}}(\mathbf{r})=\sum_{\lambda} \lambda \psi_{\lambda}(\mathbf{r})$ and the distribution of the spin electrochemical potential $\bar{\phi}_{\text{sp}}=\delta \tilde{\mu} / e$ are shown in Fig.~\ref{fig:streamlines8}. As in the opposite-contact geometry shown in Fig.~\ref{fig:streamlines6}, altermagnetism leads to nontrivial configurations of the spin current even when only an electric current is injected into the sample. At $t_1 \neq 0$, we obtained two enclosed loops at the source and the drain as well as a vortex between them. At $t_2 \neq 0$, only two enclosed loops emerge.

The characteristic values of the key quantities, i.e., the magnetic field $B_0=I/L$ and the electrochemical potential $\phi_0 = I/\left(e\tau \tilde{t}_0 \rho^{(0)}\right)$ are
\begin{eqnarray}
\label{estimate-B0}
B_0[\mu\mbox{T}] &=& 1.26~\frac{I[\mu\mbox{A}]}{L[\mu\mbox{m}]},\\
\label{estimate-phi0}
\phi_0[\mu\mbox{V}] &=& 17.0~\frac{I[\mu\mbox{A}]}{\mu[\mbox{meV}] \tau[\mbox{ps}]}.
\end{eqnarray}
For our numerical estimates, we use the material parameters listed in the main text for RuO$_2$ and KRu$_4$O$_8$. We consider an injected current similar to that in graphene, $I=0.1~\mu\mbox{A}$~\cite{Bandurin-Polini:2016}. We obtain relatively modest values of the characteristic magnetic field $B_0 \approx 25~\mbox{nT}$ and voltage $\phi_0 \approx 4.2~\mu\mbox{V}$, which is about two orders of magnitude smaller than in graphene~\cite{Torre-Polini:2015}.

\section{Fixed potential at the contacts}
\label{sec:potential}

In this section, we consider a different type of boundary condition in which we fix the electric potential and spin density at the contacts, i.e., $\bar{\phi}_{\lambda}$. As in Sec.~\ref{sec:analyt}, we focus on the ribbon geometry. We use Eq.~\eqref{eq:transport-eq-current-lambda-lin-phi} and fix $\bar{\phi}_{\lambda}(x,y=0)=\bar{\phi}_{1,\lambda}(x)$ and $\bar{\phi}_{\lambda}(x,y=L)=\bar{\phi}_{2,\lambda}(x)$. Performing the Fourier transform with respect to $x$ in Eq.~\eqref{eq:transport-eq-current-lambda-lin-phi}, we obtain the following solution for $\bar{\phi}_{\lambda, k}(y)$:
\begin{equation}
\label{eq:phi-lambda}
\bar{\phi}_{\lambda, k}(y)= \frac{1}{e^{\varkappa_{2} L} -e^{\varkappa_{1} L}} \left\{ \left[\bar{\phi}_{1,\lambda}(k) e^{\varkappa_{2} L} - \bar{\phi}_{2,\lambda}(k) \right] e^{\varkappa_{1} y} +\left[\bar{\phi}_{2,\lambda}(k) - \bar{\phi}_{1,\lambda}(k) e^{\varkappa_{1} L} \right] e^{\varkappa_{2} y} \right\}, 
\end{equation}
where $\varkappa_{1,2}$ are given in Eq.~\eqref{eq:analyt-varkappa}. The above solution, however, is not very practical since the distribution of the electrochemical potential is known only at the contacts. It is possible to rewrite such boundary conditions in terms of the fixed currents if the conductance of the contacts is provided. This leads, however, to an integral equation for $\bar{\phi}_{\lambda, k}(y)$, which is more computationally demanding and less transparent. 

 As an alternative, we use the finite element method (FEM) to solve the partial differential equation \eqref{eq:transport-eq-current-lambda-lin-phi}. We assume contacts of the finite width $2w$,
\begin{equation}
\label{eq:phi-lambda-gen-bc}
\begin{aligned}
\bar{\phi}_{\lambda} \left(x,y=0\right)&=\bar{\phi}_{1,\lambda}(x),\,\,|x|<w,\\
\bar{\phi}_{\lambda} \left(x,y=L\right)&=\bar{\phi}_{2,\lambda}(x),\,\,|x|<w,\\
\end{aligned}
\end{equation}
and set the normal component of the current to zero away from the contacts, 
\begin{equation}
\label{eq:j-lambda-gen-bc}
\begin{aligned}
j_{\lambda,y} \left(x,y=0\right)&=e \tau  \Pi_{iy}^{(0)}\nabla_{i}\bar{\phi}_{\lambda}(x,y=0)=0,\,\,|x|>w,\\
j_{\lambda,y} \left(x,y=L\right)&=e \tau  \Pi_{iy}^{(0)}\nabla_{i}\bar{\phi}_{\lambda}(x,y=L)=0,\,\,|x|>w.
\end{aligned}
\end{equation}
In addition, to compare our results with those obtained in the Keldysh approach on the lattice, we consider a rectangular sample with
$x\in \left[ -L_x, L_x \right]$ and $y \in \left[ 0 , L_y \right]$.

\begin{figure}[h]
\centering
\includegraphics[width=.99\textwidth]{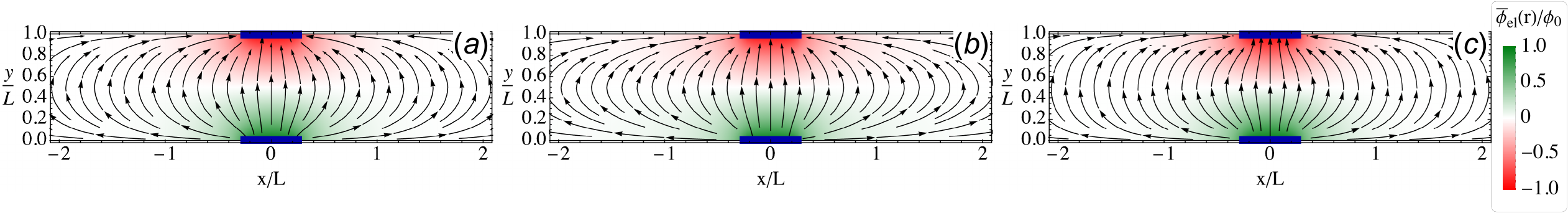}
\caption{
The electric current streamlines for: (a) $t_{1}=t_{2}=0$; (b) $t_1=t_0/2 $, $t_2=0$; (c) $t_1=0$, $t_2=t_0/2$. The electrochemical potential $\bar{\phi}_{\text{el}} \left(\mathbf{r}\right) / \phi_0$ is shown in color. The green and red colors correspond to $\bar{\phi}_{\text{el}}(\mathbf{r})>0$ and $\bar{\phi}_{\text{el}}(\mathbf{r})<0$, respectively. The blue rectangles denote the contacts. 
In all panels, we fixed $\bar{\phi}_{1,\lambda}(x)=-\bar{\phi}_{2,\lambda}(x)=\phi_{0}/2$.
}
\label{fig:streamlines-potential-el-1}
\end{figure}

\begin{figure}[h]
\centering
\includegraphics[width=.99\textwidth]{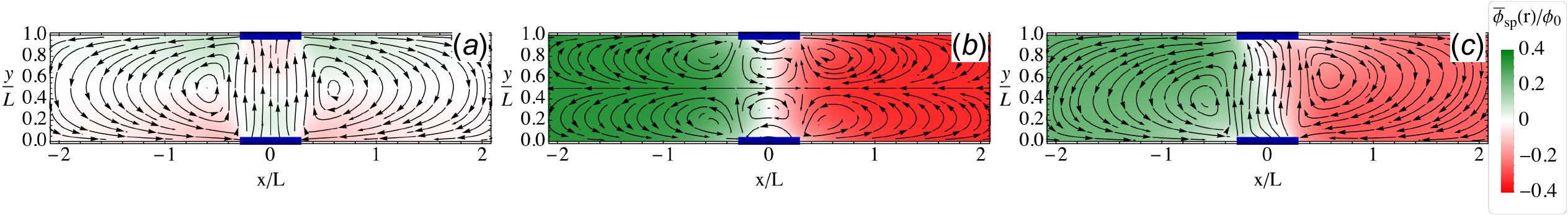}
\caption{
The spin current streamlines for: (a) $t_{1}=t_0/2$, $t_2=0$; (b) $t_1=0$, $t_2=t_0/2$; (c) $t_1=t_2=t_0/(2\sqrt{2})$. The electrochemical potential $\bar{\phi}_{\text{sp}} \left(\mathbf{r}\right) / \phi_0$ is shown in color. The green and red colors correspond to $\bar{\phi}_{\text{sp}}(\mathbf{r})>0$ and $\bar{\phi}_{\text{sp}}(\mathbf{r})<0$, respectively. The blue rectangles denote the contacts.
In all panels, we fixed $\bar{\phi}_{1,\lambda}(x)=-\bar{\phi}_{2,\lambda}(x)=\phi_{0}/2$.
}
\label{fig:streamlines-potential-sp-1}
\end{figure}

The results for $L_x=3L_y=3L$ and $\bar{\phi}_{1,\lambda}(x)=-\bar{\phi}_{2,\lambda}(x)=\phi_{0}/2$ are shown in Figs.~\ref{fig:streamlines-potential-el-1} and \ref{fig:streamlines-potential-sp-1}; we used a rectangle sample with $x\in \left[ 0, L \right]$ and $y \in \left[ -L/2 , L/2 \right]$ in the main text. 
The corresponding streamlines are similar to those discussed in the main text, cf. Fig.~2. Note that these current streamlines have the same shape as the streamlines in Fig.~\ref{fig:streamlines6}, where we fixed the electric current at the contacts. In addition, the finiteness in the $x$ direction does not qualitatively affect the streamlines.

Since the spin-flipping scattering was neglected in this section, the equations of motion and the boundary conditions are symmetric with respect to replacing the electric and spin degrees of freedom.
For example, the electric and spin currents at $\bar{\phi}_{1,\lambda}(x)=-\bar{\phi}_{2,\lambda}(x)=\lambda \phi_{0}/2$ have the same distribution as the spin and electric current at $\bar{\phi}_{1,\lambda}(x)=-\bar{\phi}_{2,\lambda}(x)=\phi_{0}/2$, see Figs.~\ref{fig:streamlines-potential-sp-1} and \ref{fig:streamlines-potential-el-1}.
The corresponding circulating electric currents induce a magnetic field, which can be measured via magnetometry techniques~\cite{Levine-Walsworth-PrinciplesTechniquesQuantum-2019, Barry-Walsworth:2019-QSM}; see also the discussion in the main text.

To find the magnetic field induced by a flowing electric current, we employ the static Maxwell equations, 
\begin{equation}
\left(\bm{\nabla} \cdot \mathbf{B} \right) = 0, \quad 
\left[\bm{\nabla} \times \mathbf{B} \right] =4 \pi \mathbf{J}_{\text{el}}.
\end{equation}
In our case of the ribbon geometry, the electric current density $\mathbf{J}_{\text{el}}$ is
\begin{equation}
\mathbf{J}_{\text{el}}=\mathbf{j}_{\text{el}}\left( \mathbf{r}\right) \delta (z)  \left[ \Theta\left(L-y \right)-\Theta\left(y \right) \right].
\end{equation}
Introducing the vector potential $\mathbf{A}\left( \mathbf{r} , z\right) $ and using the Coulomb gauge $\bm{\nabla} \cdot \mathbf{A}\left( \mathbf{r} , z\right)=0$, we obtain
\begin{equation}
-\nabla^2\mathbf{A} =4 \pi \mathbf{j}_{\text{el}}  \left( \mathbf{r} \right) \delta (z) \left[ \Theta\left(L-y \right)-\Theta\left(y \right) \right].
\end{equation}
The solution for $\mathbf{A}\left( \mathbf{r} , z\right) $ reads as
\begin{equation}
\mathbf{A}\left( \mathbf{r} , z\right) = \int_{-\infty}^{+\infty} dx^{\prime} \int_{0}^{L} dy^{\prime} \int_{-\infty}^{+\infty} dz^{\prime} \frac{\mathbf{j}_{\text{el}}  \left( \mathbf{r}^{\prime}\right)\delta (z^{\prime})}{\sqrt{ \left|\mathbf{r}-\mathbf{r}^{\prime} \right|^2+\left(z-z^{\prime} \right)^2 }}=\int_{-\infty}^{+\infty} dx^{\prime} \int_{0}^{L} dy^{\prime} \frac{\mathbf{j}_{\text{el}}  \left( \mathbf{r}^{\prime}\right)}{\sqrt{ \left|\mathbf{r}-\mathbf{r}^{\prime} \right|^2+z^2 }}.
\end{equation}
Hence, we obtain the following expression for the magnetic field:
\begin{equation}
\label{eq:magnetic-field}
\mathbf{B}\left( \mathbf{r} , z\right) =\int_{-\infty}^{+\infty} dx^{\prime} \int_{0}^{L} dy^{\prime} \frac{z  j_{\text{el}, y}\left( \mathbf{r}^{\prime}\right) \hat{\mathbf{e}}_{1}- z j_{\text{el}, x} \left( \mathbf{r}^{\prime}\right) \hat{\mathbf{e}}_{2}  +\left[ \left( y-y^{\prime} \right) j_{\text{el}, x} \left( \mathbf{r}^{\prime}\right) - \left( x-x^{\prime} \right) j_{\text{el}, y} \left( \mathbf{r}^{\prime}\right) \right] \hat{\mathbf{e}}_{3}}{\left( \left|\mathbf{r}-\mathbf{r}^{\prime}\right|^2+z^2 \right)^{3/2}},
\end{equation}
where $\hat{\mathbf{e}}_i$ is the unit vector in the $i$th direction. We show the distribution of the magnetic field in Fig.~3 in the main text.

\section{Role of spin-flip processes}
\label{sec:spin-flip}

To explore the role of spin-flip processes in current flows, we add the following term in the collision integral $I_{\text{col}}\left\{ f_{\lambda}\right\}$:
\begin{equation}
\label{sm-sp-ci}
I_{\text{sp}}\left\{ f_{\lambda} \right\} =-\frac{f_{\lambda}- \langle f_{-\lambda}\rangle }{\tau_{\text{sp}}},
\end{equation}
where $\tau_{\text{sp}}$ is the spin-flip relaxation time and $\langle f_{-\lambda}\rangle$ is averaged over the Fermi surface distribution function of the opposite spin projection. In this case, we have the following analogs of Eqs.~\eqref{eq:transport-eq-rho-lambda} and \eqref{eq:transport-eq-current-lambda}:
\begin{eqnarray}
\label{eq:rho-eq-sf}
&&\frac{\partial \rho_\lambda}{\partial t}+\left(\bm{\nabla} \cdot  \mathbf{j}_\lambda \right)=-\frac{\rho_{\lambda}-\rho_{-\lambda}}{\tau_{\text{sp}}},\\
\label{eq:j-eq-sf}
&&\frac{\partial j_{\lambda,j}}{\partial t} + \nabla_{i} \tilde{\Pi}_{ij} +e\Pi_{ij}E_i  = - \frac{j_{\lambda,j}}{\tau}.
\end{eqnarray}
Here, $1/\tau = 1/\tau_{\rm el} + 1/\tau_{\rm sp}$ is the total scattering time.

Similarly to Eq.~\eqref{eq:transport-eq-current-lambda-lin}, we obtain the following equation for the electrochemical potential:
\begin{equation}
\label{eq:main-phi-dbc-sp}
\nabla_{j} \left( \Pi_{ij}^{(0)}\nabla_{i}\bar{\phi}_{\lambda} \right)= -\frac{\nu_{\lambda}^{(0)}\bar{\phi}_{\lambda} -\nu_{-\lambda}^{(0)}\bar{\phi}_{-\lambda}}{2 \tau \tau_{\text{sp}}},
\end{equation}
where we used $\rho_{\lambda}(t,\mathbf{r}) = \rho_{\lambda}^{(0)}+ e \nu_{\lambda}^{(0)} \delta \mu_{\lambda}(t,\mathbf{r})$. 

Assuming $\mu_{+}^{(0)}=\mu_{-}^{(0)}$, Eq.~\eqref{eq:main-phi-dbc-sp} can be rewritten as
\begin{equation}
\label{eq:phi-lambda-eq-sp}
t_{0} \nabla^2 \bar{\phi}_{\lambda} + \lambda \left[T_{xx} \left( \nabla_{x}^2 - \nabla_{y}^2\right)  +2 \lambda T_{xy} \nabla_{x} \nabla_{y}  \right]\bar{\phi}_{\lambda} = -\lambda \nu^{(0)} \frac{\bar{\phi}_{\rm sp}}{2 \tau \tau_{\text{sp}}},
\end{equation}
where $\nu_{+}^{(0)}=\nu_{-}^{(0)}=\nu^{(0)}/2$ and $\rho_{+}^{(0)}=\rho_{-}^{(0)}=\rho^{(0)}/2$. Equation \eqref{eq:phi-lambda-eq-sp} is, in fact, a coupled system of two equations,
\begin{eqnarray}
\label{eq:phi-lambda-eq-sp-1}
&&t_{0} \nabla^2 \bar{\phi}_{\text{el}} + T_{xx} \left( \nabla_{x}^2 - \nabla_{y}^2 \right)  \bar{\phi}_{\text{sp}} +2 T_{xy} \nabla_{x} \nabla_{y}  \bar{\phi}_{\text{sp}} =0,\\
\label{eq:phi-lambda-eq-sp-2}
&&t_{0} \nabla^2 \bar{\phi}_{\text{sp}} + T_{xx} \left( \nabla_{x}^2 - \nabla_{y}^2 \right)  \bar{\phi}_{\text{el}} +2 T_{xy} \nabla_{x} \nabla_{y}  \bar{\phi}_{\text{el}}  = -\frac{e \nu^{(0)}}{\tau \tau_{\text{sp}} \rho^{(0)}} \bar{\phi}_{\text{sp}}.
\end{eqnarray}

\begin{figure}[h]
\centering
\includegraphics[width=.99\textwidth]{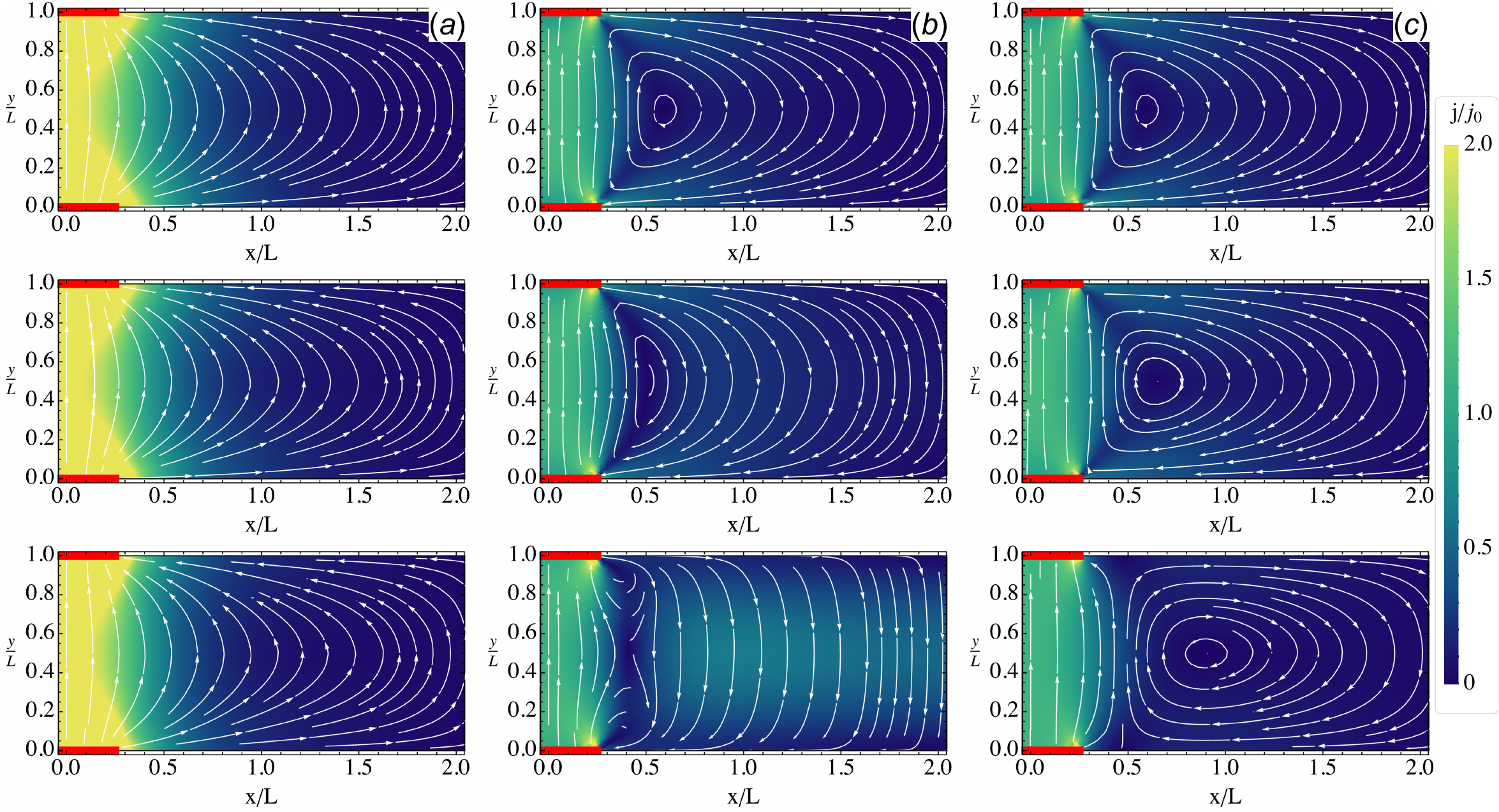}
\caption{
The current stream lines at $t_1=t_0/2 $, $t_2=0$ and (from top to bottom) $\alpha_{\text{sp}}=0$, $\alpha_{\text{sp}}=2$, $\alpha_{\text{sp}}=4$ for: (a) electric current with $\bar{\phi}_{1,\lambda}(x)=-\bar{\phi}_{2,\lambda}(x)= \phi_{0}/2$; (b) spin current with $\bar{\phi}_{1,\lambda}(x)=-\bar{\phi}_{2,\lambda}(x)= \phi_{0}/2$; (c) electric current with $\bar{\phi}_{1,\lambda}(x)=-\bar{\phi}_{2,\lambda}(x)=\lambda \phi_{0}/2$. The currents are normalized by $j_{0} = 2e\tau t_0 \, \mbox{max}{\left\{\left|\rho_{\lambda}^{(0)} \left(\bar{\phi}_{2,\lambda}-\bar{\phi}_{1,\lambda}\right) \right|\right\}}/L$ where maximal value is taken with respect to the spin projections, and we fixed $\rho^{(0)}_{\uparrow}=\rho^{(0)}_{\downarrow}=\rho^{(0)}/2$. 
The red rectangles denote the contacts. In all panels, we set $w=L/4$ and $L_x=3L_y=3L$.
}
\label{fig:streamlines-potential-el-1-spfl}
\end{figure}

By using the same geometry as in Sec.~\ref{sec:potential}, we obtain the electric and spin current distributions together with the corresponding electrochemical potential, see Figs.~\ref{fig:streamlines-potential-el-1-spfl}(a) and \ref{fig:streamlines-potential-el-1-spfl}(b) for several values of the spin-flip parameter $\alpha_{\text{sp}}= e \nu^{(0)} L^2 /(\tau \tau_{\text{sp}} t_0 \rho^{(0)})$.

The magnitude of the spin-flip parameter can be estimated as follows
\begin{equation}
\alpha_{\text{sp}} =\frac{e \nu^{(0)} L^2 }{\tau \tau_{\text{sp}} t_0 \rho^{(0)} } \lesssim \frac{L^2}{\tau^2 \mu t_0} \simeq 4.3\times10^{4} 
\frac{L^2[\mu\mbox{m}]}{\mu [\mbox{meV}] \left(a[\mbox{\AA}]\right)^2 (t_0a^{-2})\left[\mbox{eV}\right] \tau^2[\mbox{ps}]},
\end{equation}
where we assumed $\tau_{\text{sp}} \gtrsim \tau$ in the second expression. For $\tau\approx 1~\mbox{ps}$, $L = 1 ~\mu\mbox{m}$ and $t_0a^{-2}=2.5~\mbox{eV}$, $a=4.5~\mbox{\AA}$, and $\mu=0.4~\mbox{eV}$ as in RuO$_2$~\cite{Ahn-Kunes:2019, Smejkal-Jungwirth-GiantTunnelingMagnetoresistance-2022}, we estimate $\alpha_{\text{sp}} \lesssim 2.14$.

As one can see from Fig.~\ref{fig:streamlines-potential-el-1-spfl}(a), spin-flipping scattering does not qualitatively affect the electric current streamlines. For the spin current streamlines shown in Fig.~\ref{fig:streamlines-potential-el-1-spfl}(b), we obtain the decay of the vortices if only a voltage difference is applied to the contacts.

Since we no longer have the symmetry by which we can interchange the electric and spin currents if the spin imbalance is maintained at the contacts instead of the voltage difference, let us calculate the current streamlines and the corresponding electrochemical potential at $\bar{\phi}_{1,\lambda}(x)=-\bar{\phi}_{2,\lambda}(x)=\lambda \phi_{0}/2$. We show the corresponding results in Fig.~\ref{fig:streamlines-potential-el-1-spfl}(c). The spin-flipping scattering broadens the vortices of the electric current. In addition, the absolute value of the electric current decays with $\alpha_{\rm sp}$.

\begin{figure}[h]
\centering
\includegraphics[width=.99\textwidth]{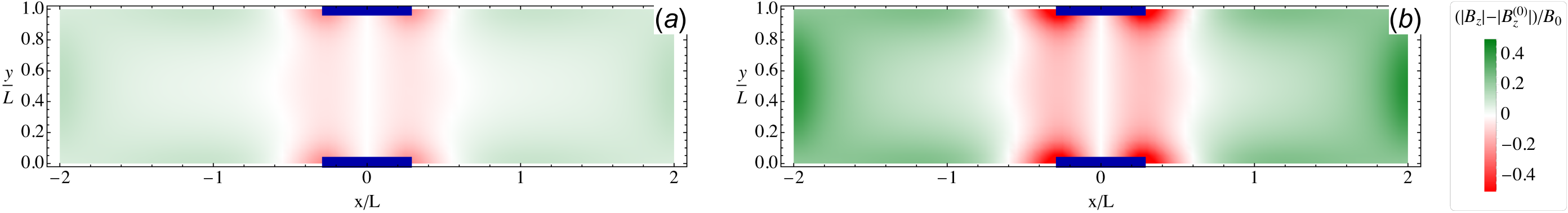}
\caption{
The difference of the magnetic fields $B_{z}(\mathbf{r})$ at  $\alpha_{\text{sp}}\neq 0$ and $B_{z}^{(0)}(\mathbf{r})$ at  $\alpha_{\text{sp}}=0$ for $t_1=t_0/2 $, $t_2=0$ and: (a) 
$\alpha_{\text{sp}}=1$; (b) $\alpha_{\text{sp}}=2$. Here $B_0=e \tau \rho^{(0)} t_0 \phi_0 / L$. In all panels, we set $w=L/4$, $L_x=3L_y=3L$, and $\bar{\phi}_{1,\lambda}(x)=-\bar{\phi}_{2,\lambda}(x)=\phi_{0}/2$. The blue rectangles denote the contacts. 
}
\label{fig:streamlines-potential-el-1-spfl-magnetic}
\end{figure}

The role of the spin-flipping processes on the electric current flows can be quantified by the magnetic field, which we show in Fig.~\ref{fig:streamlines-potential-el-1-spfl-magnetic}. As expected, the maximum value of $B_z$, which is determined by the maximal value of the electric current, decreases with increasing $\alpha_{\text{sp}}$.

\section{Details of lattice model and Keldysh technique}
\label{sec:Keldysh}

In this section, we present additional details and results of the Keldysh approach. To highlight the difference brought by altermagnetic spin-polarized bands, we also consider a ferromagnet. The latter is described by the same model as the altermagnet, %see the Appendix for the model, 
albeit with $t_m=0$ and additional spin-dependent on-site potential $\varepsilon_m \sigma_z$.

For the sake of completeness, we repeat the key details of the model presented in the main text here as well. A $d_{xy}$-altermagnet is modeled by the following tight-binding Hamiltonian:
\begin{equation}
    H_S = \sum_{i \sigma} \varepsilon_i c^\dagger_{i\sigma}c_{i \sigma}
    +\varepsilon_m \sum_{i \sigma} c^\dagger_{i\sigma} \sigma_zc_{i \sigma}
    + \sum_{ij\sigma \sigma'} c_{i\sigma}^\dagger t_{ij}^{\sigma \sigma'}c_{j\sigma'},
\end{equation}
where $c_{i\sigma}$ ($c_{i\sigma}^{\dag}$) is the annihilation (creation) operator of electrons with the spin $\sigma$ at the site $i$. The onsite potential and its spin-dependent counterpart are $\varepsilon_i$ and $\varepsilon_m \sigma_z$, respectively. The spin-dependent hopping terms are
\begin{equation}
    \hat{t}_{ij} = \begin{cases}
    -t \mathbb I, \quad (j = i \pm e_x \; \mathrm{or} \; j = i \pm e_y) \\
    -t_m \sigma_z, \quad (j = i \pm (e_x + e_y) \\
    t_m \sigma_z, \quad (j = i \pm(e_x - e_y),\end{cases}
\end{equation}
where $e_i$ is the unit vector in the $i$-th direction and $t_m$ corresponds to the strength of the spin-dependent diagonal hopping. The leads and connection between the leads and sample region are modeled as metals with $\varepsilon_i = 0$ and $\varepsilon_m = 0$, and we only consider nearest neighbor hopping with $t_{ij}^{\mathrm{leads}} = -t \mathbb I$ and $t_{ij}^{\mathrm{interface}} = -t \mathbb I$, respectively. The constrictions between the leads and the sample are modeled by a very high on-site potential $\varepsilon_{\mathrm{edge}} = 10^5\,t$ on the edges of the sample, except for a region of ten sites where the on-site potential is zero.

We define a bond spin-current operator representing the flow of spin $S_k$ from the site $i$ to $j$ as
\begin{equation}
    J_{ij}^{S_k} \equiv \frac{1}{4i}\sum_{\alpha \beta}\left(c_{j\beta}^\dagger \{\sigma_k, t_{ji} \}_{\beta \alpha} c_{i \alpha} - \mathrm{H.c.}\right),
\end{equation}
whose steady-state non-equilibrium statistical average in the Keldysh formalism is given by
\begin{equation}
\label{SM-Keldysh-J-fin}
    \langle J^{S_k}_{ij} \rangle = \frac{1}{2\pi i} \int_{-\infty}^\infty dE \, \mathrm{Tr} (J^{S_k}_{ij} G^<_{ij} ),
\end{equation}
where the trace is performed over the spin degrees of freedom. 

The lesser Green's function entering Eq.~\eqref{SM-Keldysh-J-fin} is given by the Keldysh equation for mesoscopic transport
\begin{equation}
\label{em-G<}
    G^< = G^R \Sigma^< G^A,
\end{equation}
where $G^R$ and $G^A = (G^R)^\dagger$ are the retarded and advanced Green's functions and $\Sigma^<$ is the lesser self-energy. Assuming that the electrons supplied by the thermalized leads are scattered ballistically through the sample, the expression in Eq.~\eqref{em-G<} can be calculated exactly by replacing the effects of the semi-infinite leads via the self-energy terms
\begin{equation}
    G^R(E) = \frac{1}{E\mathbb I - H_S - \sum_i \Sigma_i(E - eV_i)},
\end{equation}
and
\begin{equation}
    \Sigma^<(E) = i \sum_i \Gamma_i(E - eV_i) f_i(E - eV_i),
\end{equation}
where $H_S$ is the Hamiltonian of the finite sample region, $\Sigma_i = H_C \,g_L \, H_C^\dagger$ are the exactly calculable self-energy terms given in terms of the connection Hamiltonian $H_C$ and the bare retarded Green's function of the leads $g_L$, $\Gamma_i \equiv i(\Sigma_i - \Sigma_i^\dagger)$, $eV_i$ is the voltage applied to the lead $i$, and $f_i$ is the corresponding Fermi-Dirac distribution.

To include disorder in the altermagnet, we uniformly assign a fraction $n_I = 0.2$ of the sites with a local potential of strength $\varepsilon_{\rm imp} = 10\,t_m$ simulating pointlike impurities. The disorder average is taken with respect to $100$ random configurations of impurities.

\begin{figure}[t]
\centering
\subfigure[\label{fig:SM-Keldysh-a}]{\includegraphics[width=0.45\textwidth]{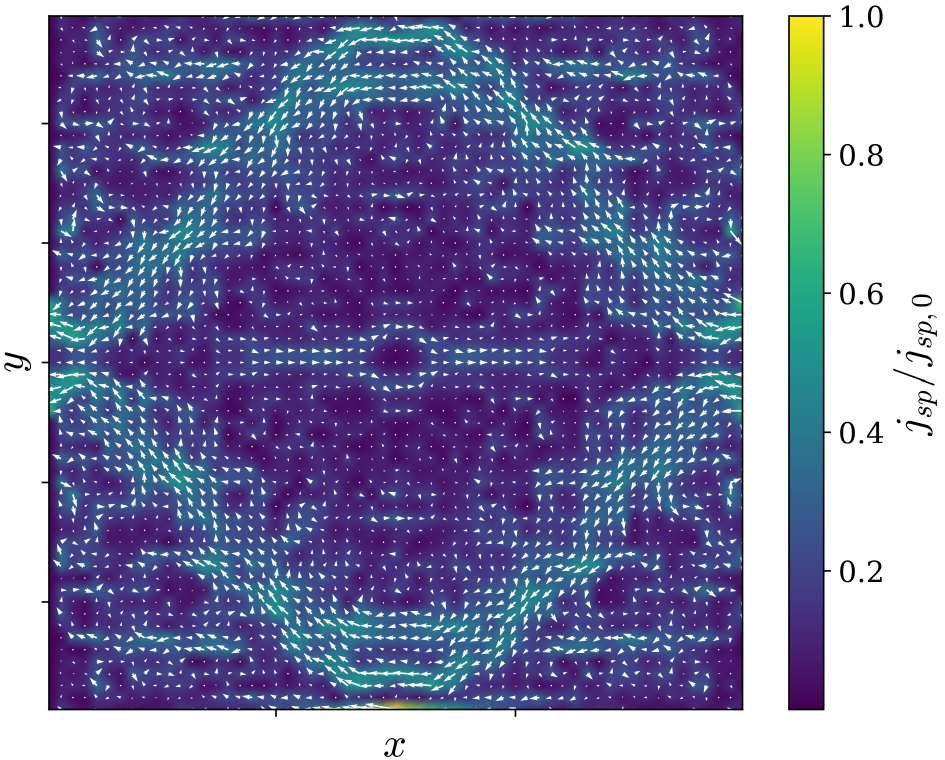}}
\subfigure[\label{fig:SM-Keldysh-b}]{\includegraphics[width=0.45\textwidth]{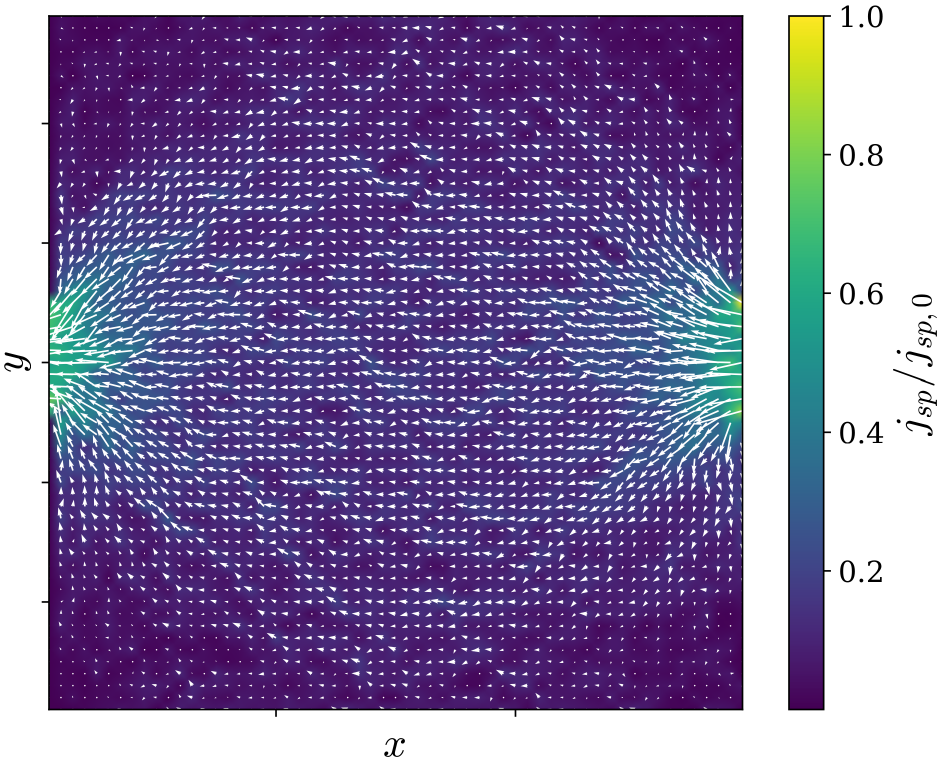}}\\
\subfigure[\label{fig:SM-Keldysh-c}]{\includegraphics[width=0.45\textwidth]{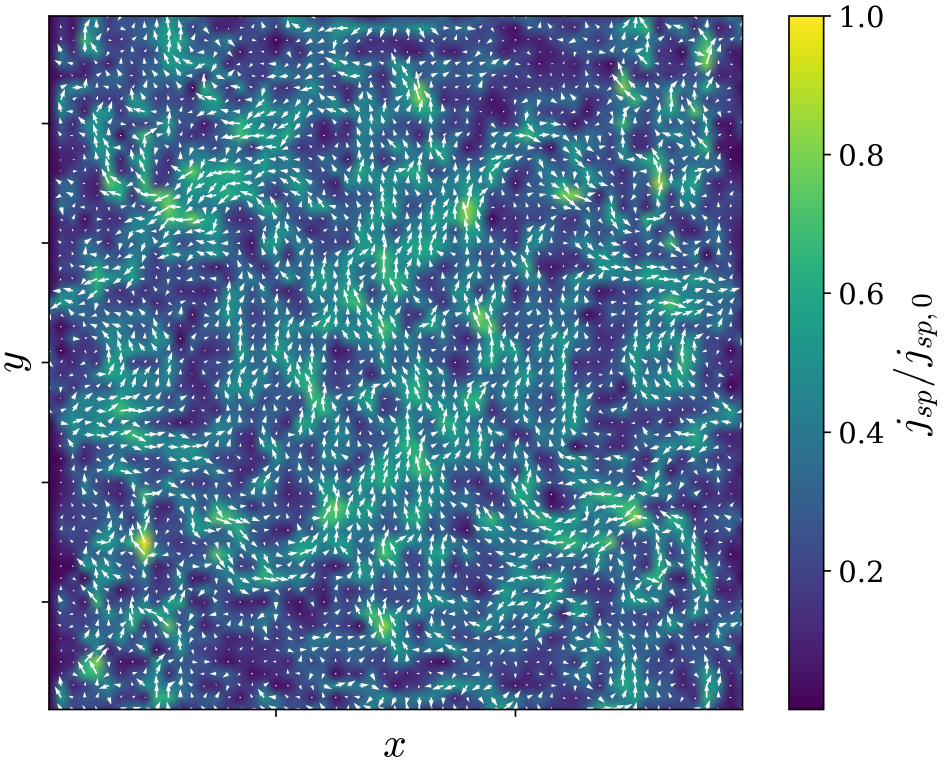}}
\subfigure[\label{fig:SM-Keldysh-d}]{\includegraphics[width=0.45\textwidth]{60x60_0.3_AM_diffusive_spin_current_cropped.pdf}}\\
\caption{
Spin current distribution in the ballistic (left column) and diffusive (right column) transport regimes in ferromagnet (top row) and altermagnet (bottom row) obtained via the Keldysh formalism. The contacts of the width $w=10\,a$ are at $x=0,L_x$ surfaces. We use $60\times 60$ lattice with the following parameters: $t_m = 0.3\,t$ (altermagnet), $\varepsilon_m = t$, $eV = 0.2\,t$, and $\mu = -2\,t$. The spin current is normalized to the maximum spin current $j_{sp, 0}$.
}
\label{fig:SM-Keldysh}
\end{figure}

We compare spin current distributions in ballistic (clean) and diffusive (averaged over several disorder configurations) transport regimes for an altermagnet and a ferromagnet in Fig.~\ref{fig:SM-Keldysh}. As one can see from Fig.~\ref{fig:SM-Keldysh-a}, the spatial distribution of the spin current is nontrivial even in ferromagnets. It strongly depends on the shape of the Fermi surface and may even show vortical structures. The spin current in clean altermagnets shown in Fig.~\ref{fig:SM-Keldysh-c} acquires an even more complicated pattern. The introduction of disorder smoothes nontrivial features, leading to a spin current distribution in a ferromagnet similar to that for the electric current, see Fig.~\ref{fig:SM-Keldysh-b}. On the other hand, as we discussed in the main text, nontrivial vortical features in the spin current of altermagnets remain even in the diffusive regime, see Fig.~\ref{fig:SM-Keldysh-d}. This observation supports the uniqueness of the spin current vortices in disordered altermagnets.

\bibliography{library-short}